\documentclass[journal=jacsat,manuscript=article]{achemso}

\usepackage[version=3]{mhchem} 
\usepackage{xcolor,ulem}
\usepackage{xr}
\author{Denis C\'{e}olin}
\affiliation[Soleil]{Synchrotron SOLEIL, L’Orme des Merisiers, Saint-Aubin, BP 48, F-91192 Gif-sur-Yvette Cedex, France}
\altaffiliation{These authors contributed equally to this work}

\author{Tsveta Miteva}
\affiliation[LCPMR]{Laboratoire de Chimie Physique-Matière et Rayonnement, CNRS, Sorbonne Université, F-75005 Paris Cedex 05, France}
\altaffiliation{These authors contributed equally to this work}
\email{tsveta.miteva@sorbonne-universite.fr}
\author{Jean-Pascal Rueff}
\affiliation[Soleil]{Synchrotron SOLEIL, L’Orme des Merisiers, Saint-Aubin, BP 48, F-91192 Gif-sur-Yvette Cedex, France}
\alsoaffiliation[LCPMR]{Laboratoire de Chimie Physique-Matière et Rayonnement, CNRS, Sorbonne Université,
F-75005 Paris Cedex 05, France}
\author{R\'{e}mi Dupuy}
\affiliation[LCPMR]{Laboratoire de Chimie Physique-Matière et Rayonnement, CNRS, Sorbonne Université,
F-75005 Paris Cedex 05, France}
\author{Thanit Saisopa}
\affiliation[Thai]{Department of Applied Physics, Faculty of Sciences and Liberal Arts, 
Rajamangala University of Technology Isan, Nakhon Ratchasima 30000, Thailand}
\author{Yuttakarn Rattanachai}
\affiliation[Thai]{Department of Applied Physics, Faculty of Sciences and Liberal Arts, 
Rajamangala University of Technology Isan, Nakhon Ratchasima 30000, Thailand}
\author{Gunnar \"{O}hrwall}
\affiliation{MAX IV Laboratory, Lund University, Box 118, SE-22100 Lund, Sweden}
\author{St\'{e}phane Carniato}
\affiliation[LCPMR]{Laboratoire de Chimie Physique-Matière et Rayonnement, CNRS, Sorbonne Université,
F-75005 Paris Cedex 05, France}
\author{Ralph P\"{u}ttner}
\affiliation{Fachbereich Physik, Freie Universit\"{a}t Berlin, Arnimallee 14, D-14195 Berlin, Germany}
\altaffiliation{These authors contributed equally to this work}

\title{Ultrafast Charge-Transfer and Auger Decay Processes in Aqueous CaCl$_2$ Solution: Insights from  Core-Level Spectroscopy}

\keywords{American Chemical Society, \LaTeX}

\begin{document}

\begin{tocentry}
\includegraphics[scale=0.33]{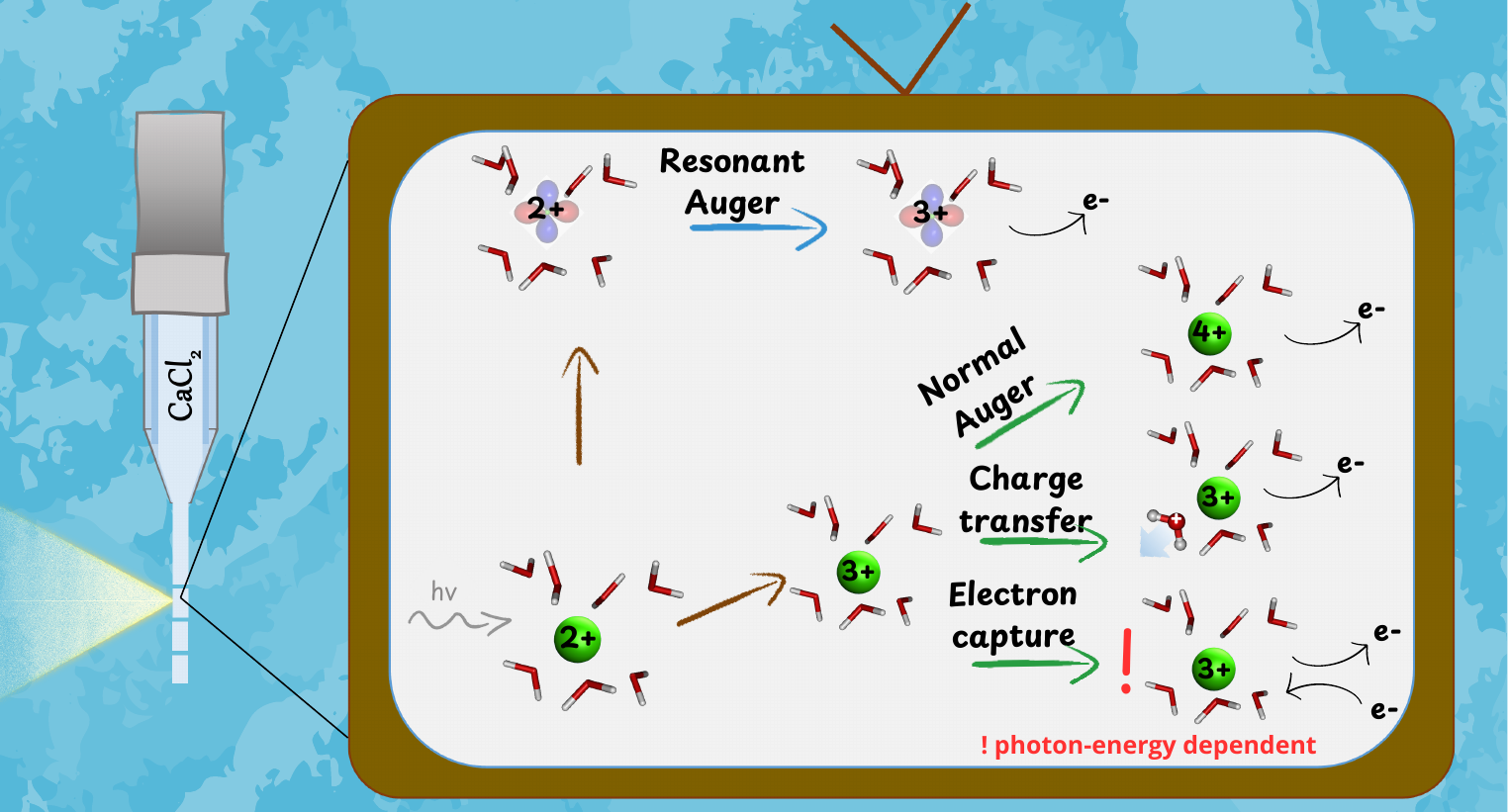}



separate page at the end of the document.

\end{tocentry}

\begin{abstract}
  Understanding the interaction between metal ions and their aqueous environment is fundamental in many areas of chemistry, biology, and environmental science. In this study, we investigate the electronic structure of hydrated calcium ions, focusing on how water molecules influence the metal ion’s behavior. We employed advanced X-ray techniques, including X-ray absorption, photoelectron, and Auger spectroscopies, combined with high-level quantum chemical calculations. Our analysis reveals that, alongside normal Auger decay, distinct ultrafast charge transfer processes occur between the calcium ion and surrounding water molecules, underscoring the complex nature of metal-solvent interactions. Two primary mechanisms were identified. The first one involves electron transfer from water to the calcium ion. The second mechanism depends on the photon energy and is tentatively attributed to the decay of photoelectron satellites, the capture of free solvated electrons or electrons from a Cl$^{-}$ ion in the second solvation shell. Additionally, we observed significant shifts in electron energies due to post-collision interactions and interpreted the Ca 1s$^{-1}$ photoelectron satellites mainly as originating from inelastic photoelectron scattering (IPES). These findings provide deeper insights into the electronic properties of hydrated metal ions, with potential implications for fields such as catalysis and biochemistry, where metal ions play a crucial role.
\end{abstract}

\section{Introduction}\label{sec:intro}
Calcium contributes up to 2\% to the weight of the human body. Almost all of this calcium can be found in the bones or the teeth. However, a very small amount can also be found in the extracellular fluid (e.g.\ blood plasma). Of this portion, roughly half exists as free, hydrated calcium ions Ca$^{2+}$ (Ref. \citep{baird11:696}), making the interaction between Ca$^{2+}$ and water an intriguing topic of study. These interactions can be effectively investigated by dissolving calcium salts in water and analyzing them with core-level spectroscopic methods, which enable simultaneous probing of both the metal ion’s electronic structure and its interactions with the solvent\citep{ceolin17:263003,dupuy24:6926}. Metal-solvent interactions modify the electronic structure of the metal ion, often resulting in observable charge transfer (CT) processes.

Charge transfer (CT) processes at the atomic scale have been subjects of investigations for decades, attracting a broad community of researchers interested in both the fundamental principles and the potential applications of these mechanisms (cf e.g.\ Ref. \citep{woerner17:061508}). Collecting information for a precise description of these processes and their timescale is essential. One way for probing the dynamics of such processes is the use of pump probe lasers, which can highlight, for example, intermolecular electron migration occurring most commonly in the femtosecond (fs) regime. Another approach for probing these mechanisms is Auger electron spectroscopy using the core-hole-clock method \citep{bjorneholm92:1892,foehlisch05:373}. This requires rather high-energy photons to get access to specific inner shells with lifetimes on the fs timescale. The evolution of the system during the lifetime of the core-hole state can be extracted by detecting the electron emitted in the decay. This technique is frequently employed at synchrotron radiation facilities.  

In this context, numerous experiments have been performed aiming at quantifying ultrafast charge transfers between various types of adsorbates and substrates \citep{foehlisch05:373,bruhwiler02:703,wurth00:141}. In the case of an adsorbate inner-shell excitation, the corresponding resonance (leading to a core-excited state of the adsorbate) and the continuum states (i.e. core-ionized states of the substrate) can overlap, leading to a possible charge transfer for the adsorbate to the substrate and, therefore, to a competition between resonant and normal Auger decays. The spectral signature of this mechanism is a contribution of the adsorbate normal Auger lines below the corresponding ionization threshold, meaning that the initially promoted core-electron migrates to the substrate before the electronic decay takes place.

This mechanism extends to simple electrolyte solutions, as demonstrated for instance on KCl dissolved in water \citep{miteva18:4457}. Here, the resonant excitation leads to a competition between the Cl$_{aq}^{-}$(1s$^{-1}$4p) $\rightarrow$ Cl$_{aq}^{0}$(2p$^{-2}$4p)  resonant Auger decay and the Cl$_{aq}^0$(1s$^{-1}$) $\rightarrow$ Cl$_{aq}^{+}$(2p$^{-2}$) normal Auger decay; in the latter case the 4p electron is transferred to the surrounding water within the lifetime of the core-hole, leaving  Cl in a neutral charge state prior to the Auger decay. Note that for reasons of simplicity, we give the charge state of each electronic state of the atoms or ions. In addition, the subscript $aq$ refers to aqueous solution.

Beside the mechanism of electron delocalization, we also found the opposite mechanism: the core-ionized species receives an electron from a neighboring water molecule at the time of the electronic relaxation, similarly to the case of KCl, see Ref. \citep{ceolin17:263003,miteva18:4457}, and of the solid-state compounds considered in Ref.\citep{nishikida78:49}.
 More precisely, the ionization of the K-shells of the potassium and chloride ions is followed mostly by an Auger relaxation leading to final ionic states with two holes in their 2p shells. However, the two corresponding KL$_{2,3}$L$_{2,3}$ Auger spectra present a significant difference: solely for K$_{aq}^{+}$, an intense structure can be found on the low kinetic-energy side of the 
$^1$D main peak. This peak is associated with a CT process accompanying the Auger decay, which competes with the   K$_{aq}^{2+}$(1s$^{-1}$) $\rightarrow$ K$_{aq}^{3+}$(2p$^{-2}$) decay without charge migration. In this specific case, the transfer of the electron happens between an occupied water orbital and an unoccupied potassium 3d orbital. Since K$_{aq}^{+}$ and Cl$_{aq}^{-}$  are isoelectronic the absence of an evident corresponding peak in the Cl$_{aq}^{-}$ spectrum is an indication that the initial charge of the species plays an important role. Calculations also highlighted the essential role of the distance between the electron donor and the electron acceptor, as well as the contribution of the relaxation energy of the ion valence orbitals after the core-hole formation.

Given these findings, we extended this study to an aqueous solution of CaCl$_2$, focusing on the distinct electronic structure and charge transfer mechanisms of Ca$^{2+}$ ions. In contrast to K$^{+}$, the Ca$^{2+}$ ion has a shorter Ca–water distance and carries a 2+ charge, both of which are expected to significantly influence charge transfer processes. These factors make Ca$^{2+}$ an ideal candidate for investigating how the charge state of the metal ion and its local environment influence charge transfer in solution. We conducted electron emission experiments on Ca$^{2+}$ and analyzed the structures in the KL$_{2,3}$L$_{2,3}$ Auger spectra of the Ca$_{aq}^{3+}$(1s$^{-1}$)
state, both near and far above the ionization threshold, to better understand these mechanisms.

Previous studies on the Ca$_{aq}^{2+}$ ion have focused on the Ca 2p photoelectron and the subsequent L$_{2,3}$M$_{2,3}$M$_{2,3}$ Auger spectrum, revealing the presence of interatomic Coulombic decay (ICD) processes between the calcium ion and surrounding water molecules \citep{pokapanich11:13430}. Moreover, in a very recent study resonant L$_{2,3}$M$_{2,3}$M$_{2,3}$ Auger spectroscopy was used by Dupuy et. al \citep{dupuy24:6926} to explore the Ca$^{2+}$ solvation shell. 
At the L-edge, non-local electronic decay processes such as ICD can occur within the same kinetic-energy range as the Auger features. However, no additional charge transfer is observed in the L-edge spectra. In the present case of 1s$^{-1}$ ionization, the Auger spectra are dominated by charge transfer processes involving the solvent. Non-local relaxation processes such as ICD and Electron Transfer Mediated decay are not observed within the kinetic energy range of normal Auger decay. 

As we will demonstrate in this study, the 1s$^{-1}$ X-ray absorption and Auger spectra of the Ca$_{aq}^{2+}$ ion differ significantly from those of the K$_{aq}^{+}$ ion \citep{ceolin17:263003,miteva18:4457}. Specifically, the X-ray absorption spectrum (XAS) of K$_{aq}^{+}$ shows two distinct resonances below the ionization threshold, whereas the spectrum of Ca$_{aq}^{2+}$ displays only one. In the KL$_{2,3}$L$_{2,3}$ Auger spectrum, charge transfer (CT) states in K$_{aq}^{+}$ appear as a shoulder on the low kinetic energy side of the $^1$D main line. In contrast, for Ca$_{aq}^{2+}$, the CT states are much more pronounced, to the extent that they obscure the $^{1}$S line. 
In contrast with the large differences we observe in the KLL Auger spectra, the L$_{2,3}$M$_{2,3}$M$_{2,3}$ Auger spectra of K$_{aq}^{+}$ (Ref. \citep{pokapanich09:7264}) and Ca$_{aq}^{2+}$ (Ref. \citep{pokapanich11:13430}) were found to be quite similar. 
The main distinction is that the weak interatomic Coulombic decay (ICD) contributions observed in the Ca$_{aq}^{2+}$ spectrum are approximately five times more intense \citep{pokapanich11:13430} than those in K$_{aq}^{+}$. For a direct comparison of these spectra, see Ref. \citep{ottosson12:1}.

In first approximation the Auger spectra recorded with different photon energies higher than the corresponding ionization energy are assumed to be identical. However, there are some effects, which can cause deviations from this assumption, like post collision interaction (PCI), recoil effects induced by the photoelectron\citep{simon14:4069,ceolin19:4877}, or Auger decays of resonances embedded in the ionization continuum. The first and the last effect contribute to the present spectra and will be discussed below. Moreover, we identified experimentally photon-energy dependent Auger transitions, which can result from different processes such as a capture of a free electron, the transfer of an electron from Cl$^{-}$ in the second solvation shell into an unoccupied orbital of the metal ion, and the Auger decay of satellites in the photoelectron spectrum. The present results clearly show the benefit of measuring Auger spectra using different photon energies, and in particular in small photon-energy steps, as nowadays routinely possible at synchrotron radiation facilities.

\section{Experimental set-up and theoretical methods}\label{sec:exptheo}

The photoelectron as well as the resonant and normal Auger spectra were recorded with a setup specifically designed  for the study of liquids by electron spectroscopy. It is mounted at the HAXPES station \citep{ceolin13:188} of the GALAXIES beamline \citep{rueff2015:175} of the SOLEIL synchrotron facility, France. The electrolyte solution is injected with a high-performance liquid chromatography pump through a glass capillary with a diameter of 25 $\mu$m facing a catcher used to extract the liquid from the vacuum chamber, both placed in a differentially pumped tube. The liquid jet is set perpendicular to both the photon beam propagation axis and to the spectrometer lens axis, which is collinear with the horizontal polarization of the light. Aqueous solutions were prepared at 0.5M and 1M concentrations by dissolving pure calcium chloride dihydrate salts in de-ionized water. All the spectra shown in this article were recorded using a 0.5M solution. However, we note that the spectra for both concentrations recorded at h$\nu$ = 5 keV do not show significant differences. For the Auger spectra, the experimental resolution is given by the spectrometer resolution of about 0.6 eV, which was obtained using 500 eV pass energy and 0.5 mm slits. For the Ca$_{aq}^{2+}$ photoelectron spectrum recorded at 5 keV, the photon bandwidth amounts to about 0.6 eV, which in combination with 0.3 mm slits for the analyzer and a pass energy of 500 eV results in a total energy resolution of less than 750 meV. For the photoelectron and the Auger spectra the photon energy calibration was performed using the O 1s$^{-1}$ XPS line of liquid water, which is located at 538.1 eV binding energy \citep{thurmer21:10558}. Note that based on the present calibration the position of the Ca$_{aq}^{2+}$(1s 
$\rightarrow$ 3d) transition in the XAS is 3.6$\pm$0.3 eV below the threshold as obtained by XPS. This value agrees well with the value of 3.8 eV obtained at the Ca 2p threshold \citep{dupuy24:6926} and confirms the present position of the threshold relative to the absorption spectrum. 

The X-ray absorption spectrum (XAS) was recorded at the BL1.1 W beamline at the Synchrotron Light Research Institute of Thailand using a resolving power $\Delta$E/E = 10$^{-4}$; for more details, see Ref. \citep{saisopa2020:146984}. 

In the course of the present work we also performed calculations to interpret the X-ray photoabsorption and photoelectron spectra at the Ca K-edge as well as the KLL normal Auger spectrum. The theoretical X-ray absorption spectra were computed for the micro-solvated clusters Ca(H$_2$O)$_m^{2+}$, $m = 6-8$. This allows us to handle the fact that the first hydration shell of calcium is not static but fluctuates over time, resulting in a distribution of the number of water molecules around calcium, centered around the average coordination number. This number was evaluated by various experimental or theoretical methods to be in the range of 5.5 to 10, with an average around 8, according to the summary presented in \citep{megyes04:7261}. The structures were optimized at the DFT level of theory using the B3LYP functional. The geometry optimization was performed with the Q-Chem computational package \citep{Shao17012015} using the 6-311++G(3df,3pd) basis set. The obtained Ca-O distances range from 2.407\,\AA~([Ca(H$_{2}$O)$_6$]$^{2+}$) to 2.501\,\AA~ ([Ca(H$_{2}$O)$_{7,8}$]$^{2+}$). These values are in good agreement with other experimental and theoretical studies (2.46 \AA~\citep{jalilehvand2001:431}, 2.42 \AA \citep{dangelo04:11857,tongraar10:10876}).

The energies and transition moments of the core-excited states of isolated Ca$^{2+}$ and the 6-, 7- and 8-coordinated [Ca(H$_2$O)$_n$]$^{2+}$ ($n = 6-8$) clusters were computed with the algebraic diagrammatic construction method for the polarization propagator \citep{schirmer82:2395} within the core-valence separation approximation (CVS-ADC(2)x) \citep{cederbaum80:206,cederbaum81:1038,barth85:867} as implemented in the Q-Chem package \citep{wenzel2014:1900,wenzel2014:4583,wormit14:774,Shao17012015}. We used the 6-311++G** basis set on all atoms, with the basis set for Ca left uncontracted. In our calculations, the orbital space was divided into core and valence spaces, with the former space comprising the 1s orbital of Ca$^{2+}$, and the latter comprising the remaining occupied orbitals. The theoretical stick spectra were convolved with a Lorentzian profile with full width at half maximum (FWHM) of 0.81 eV \citep{krause79:329}. The core-excited states were analyzed by expanding the natural orbitals occupied by the excited electron (singly-occupied natural
orbitals, SONOs) $\psi_{i}$ of the microsolvated cluster in the basis of SONOs of the isolated Ca$^{2+}$ ion, $\chi_{nl}$
\begin{equation*}
    \psi_{i} = \sum_{nl} a^{i}_{nl}\chi_{nl}
\end{equation*}
where n and l stand for the principal and orbital angular momentum quantum numbers,
respectively, as described in Ref. \citep{miteva2016:16671}. The computed energies and intensities of the core-excited states are presented in Fig.\ SI1 of the supplementary information (SI).

The X-ray photoelectron spectrum was computed for a microsolvated [Ca(H$_{2}$O)]$^{2+}$ cluster at a configuration interaction singles-doubles (CISD) level of theory including single and double electronic excitations (CISD). Intensities have been estimated within the monopolar approximation in which they are computed as the square of the overlap between the (N-1) initial and final (N-1) CIS determinants. This approximation is valid at photon energies much larger than the 1s$^{-1}$ ionization threshold.

For the interpretation of the KLL Auger spectra, a micro-solvated [Ca(H$_{2}$O)]$^{2+}$ cluster was used as a prototype system. The energies of the Ca$_{aq}^{4+}$(2p$^{-2}$) double core-hole states were computed using the occupation-restricted-multiple-active-space (ORMAS) single- and double-excitation configuration interaction method \citep{ivanic03:9364} as implemented in the GAMESS(US) package \citep{schmidt93:1347} and the 6-311+G* basis set \citep{pritchard2019:4814,blaudeau97:5016} on all atoms. The molecular-orbital space was divided as follows – the Ca 1s, 2s and O 1s orbitals were kept frozen, two electrons were ionized from the Ca 2p orbitals, and double excitations were allowed from the three highest water molecular orbitals (1b$_2$, 3a$_1$, 1b$_1$) to all virtual orbitals. The remaining orbitals were kept doubly occupied in the calculation.

\section{Results and discussion}\label{sec:results}

In order to understand the origin of the different processes following K-shell ionization, in the following we consider the X-ray absorption as well as the Ca$_{aq}^{3+}$ (1s$^{-1}$)  photoelectron and the KL$_{2,3}$L$_{2,3}$ Auger spectra. The  KL$_{2,3}$L$_{2,3}$ Auger transitions were selected since they are narrower than the other KLL Auger transitions allowing for the observation of less intense structures.  

\subsection{The Ca$_{aq}^{3+}$(1s$^{-1}$) X-ray photoelectron spectrum}

\begin{figure}[h]
    \centering
    \includegraphics[width=0.5\linewidth]{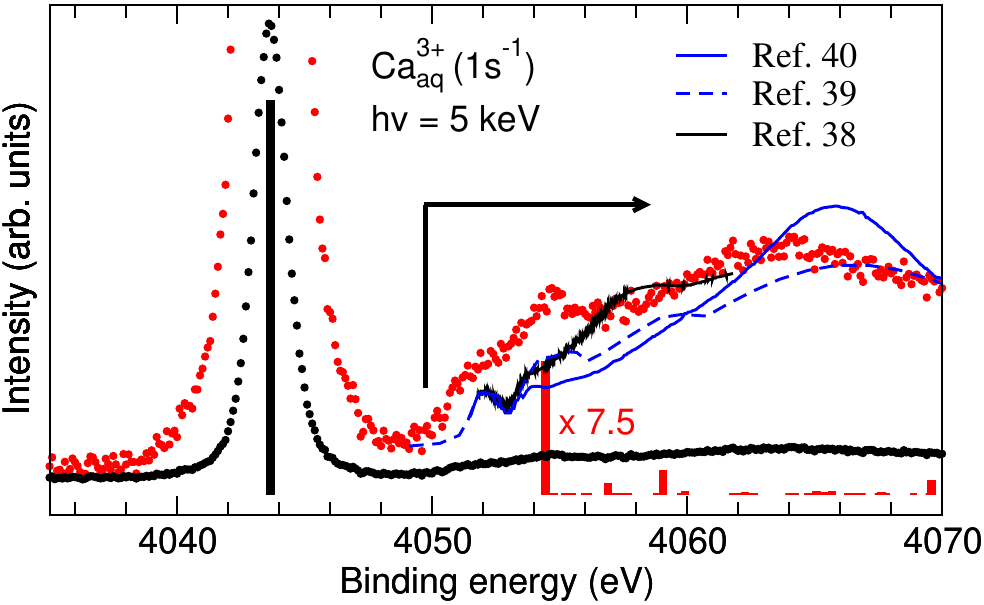}
    \caption{Ca$_{aq}^{3+}$ (1s$^{-1}$) X-ray photoelectron spectrum recorded at 5 keV photon energy (black dots). The spectrum is also scaled by a factor of 7.5 (red dots) to highlight the contribution of the satellite states above 4050 eV binding energy. The vertical bars represent the theoretical XPS stick spectrum. The red bars are multiplied by a factor of 7.5 as compared to the black bar to improve visibility. The black subspectrum below the red spectrum indicates the vacuum ultraviolet photoabsorption spectrum of liquid water, which is digitized from Ref. \citep{winter2007:094501} and aligned relative to the 1s$^{-1}$  main line. The dashed and solid blue subspectra represent theoretical energy-loss function $\eta_2(E,0)$ obtained by Dingfelder {\it et al.}\citep{dingfelder98:1} and Emfietzoglou {\it et al.}\citep{emfiezoglou03:2355}, respecitvely.}
    \label{fig:xps}
\end{figure}

In Fig.\ \ref{fig:xps} we present the experimental Ca$_{aq}^{3+}$ (1s$^{-1}$) X-ray photoelectron spectrum including the satellites as well as the theoretical results indicated by vertical bars. The red data are multiplied by a factor of 7.5 to improve the visibility of the satellite structures in the region above 4050 eV. Note that for reasons presented in the Supporting Information we avoid the misleading term
shake-up satellites, which is often used for such structures. In the following we will show that the satellite structures are predominantly due to inelastic photoelectron scattering (IPES), i.e.\ the photoelectron loses energy by exciting water molecules on its way to the detector. Because of this, the satellites should resemble electron-energy loss spectroscopy (EELS) features of water. 

As can be seen in Fig.\ \ref{fig:xps}, only partial agreement between experiment and theory is found. Beside the main line at 4043.6 eV, the calculations reproduce quite well the satellite peak around 4055 eV. This satellite is assigned based on the calculations to a process, where the photoemission of the Ca 1s electron is accompanied by a charge transfer promoting an electron of a water molecule to the Ca 3d orbital. All other satellite features present in the measured spectrum are not reproduced by the calculation.

In general, the probability of an energy transfer $E$ for a fast charged particle 
to the
surrounding medium is given in the dipole approximation by the energy-loss function $\eta_2(E,0)$; here
0 stays for the vanishing momentum transfer $q=0$ in the dipole approximation.\citep{dingfelder98:1}
To compare with the present satellite structures we show in Fig.\ \ref{fig:xps} the
theoretical energy-loss functions $\eta_2(E,0)$ obtained by Dingfelder {\it et al.}\citep{dingfelder98:1} and Emfietzoglou {\it et al.}\citep{emfiezoglou03:2355}, which are indicated by the blue dashed and solid
subspectra, respectively, and digitized from Ref.\citep{emfiezoglou03:2355}. Both spectra are calculated for electrons with a kinetic energy of 1 keV, which is very close to the present value of around 950 eV.

Moreover, the EELS features of electrons with high kinetic energy  closely resemble photoabsorption spectra. For this reason, the thin black line subspectrum below the red spectrum indicates the experimental vacuum ultraviolet photoabsorption spectrum of liquid water, digitized from Ref. \citep{winter2007:094501}. Note that all the presented subspectra exhibit a higher experimental resolution than the current spectrum. They all show reasonable agreement with each other and with the red photoelectron satellite spectrum above 4050 eV, except for the peak at 4055 eV, which is attributed to a transition directly related to the Ca 1s ionization, as discussed above. Beside the peak at 4055 eV the satellite photoelectron spectrum of Ca 1s$^{-1}$ relative to the main line agrees very well with that of K 1s$^{-1}$ and to a lesser extent to the spectrum of Cl 1s$^{-1}$, see Ref.\ \citep{ceolin17:263003}. Note that for K 1s$^{-1}$ the onset of the satellite structures approximately 6 eV above the main line was already explained with such electron-energy loss features \citep{ceolin17:263003}. From these findings we conclude that core-level photoelectron satellites of ions in water measured with photon energies well above the core-level ionization energy are generally dominated by IPES, which resembles the EELS structures of water, with some additional structures caused by the corresponding ion. This explanation is supported by the fact that electron-energy loss structures are found as very weak features in the gas-phase 1s$^{-1}$ spectrum of Ar in a gas-cell with a pressure of 10$^{-2}$ mbar or less \citep{puettner2020:052832}, i.e.\,at a target density of less than 10$^{-6}$ mol/liter. In contrast to this, the density of water in a liquid is with about 50 mol/l higher so that strong energy-loss satellites can be expected. The present interpretation of the satellite structures is fully in line with a very recent study on satellites in ambient pressure X-ray photoelectron spectroscopy using gas pressures up to 25 mbar \citep{li2024:xxxx}.

Certainly the discussed energy electron loss process is independent of the origin of the measured electron, 
i.e. such loss features are expected to be present not only in photoelectron spectra, but also in resonant and normal Auger spectra as will be presented further below.

\subsection{The Ca$_{aq}^{2+}$ K-shell X-ray photoabsorption spectrum}

\begin{figure}
    \centering
    \includegraphics[width=0.6\linewidth]{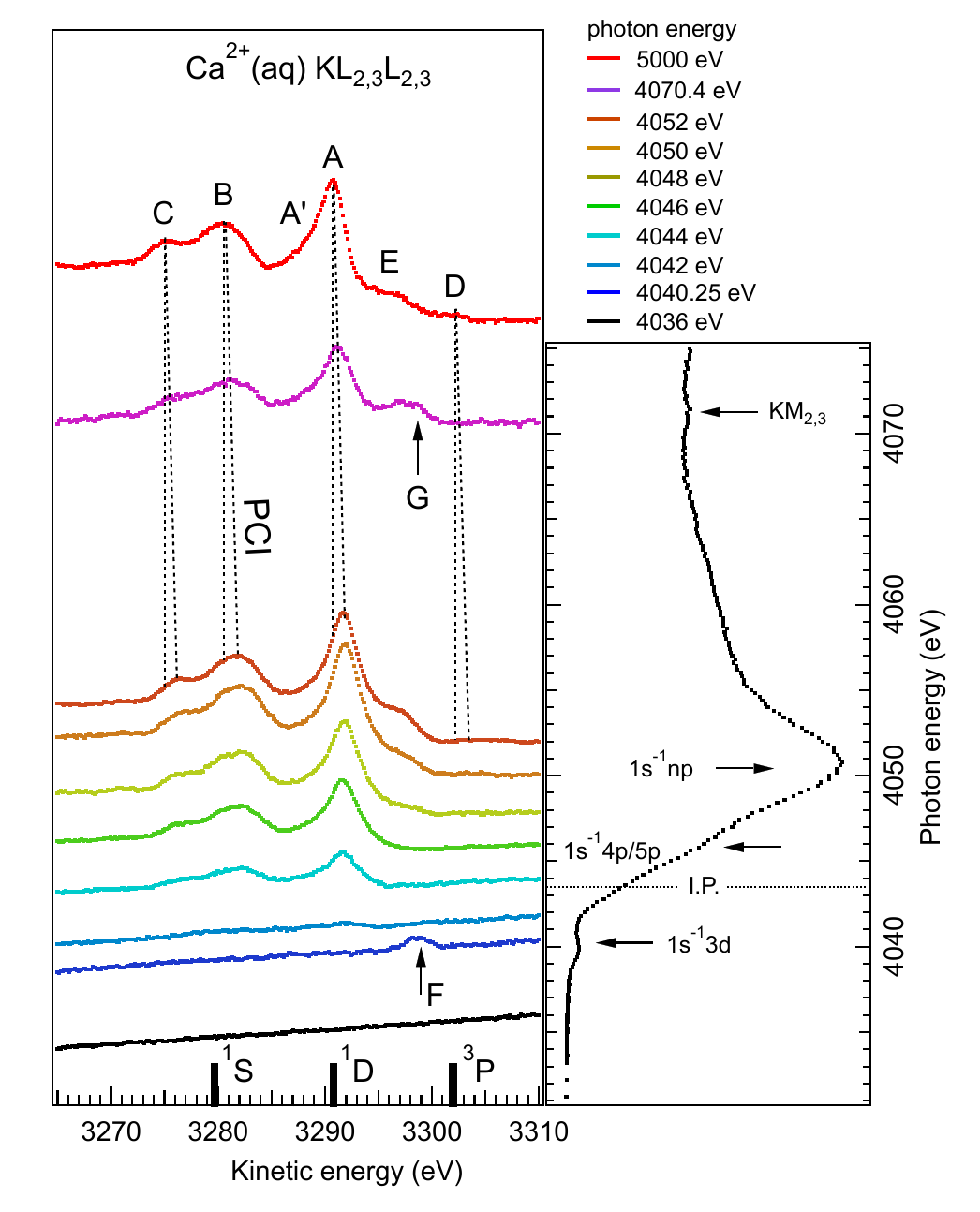}
    \caption{Right panel: X-ray absorption spectrum of Ca$_{aq}^{2+}$ recorded in the vicinity of the 1s$^{-1}$ ionization threshold indicated with I.P. Arrows indicate resonances also discussed in Ref. \citep{saisopa2020:146984}. Left panel: Resonant and normal KL$_{2,3}$L$_{2,3}$ Auger spectra recorded at various excitation energies; these energies correlate with the intensity level of each spectrum at the high-kinetic energy side and the photon-energy scale of the absorption spectrum. The off-resonance contribution is taken at a photon energy of 4036 eV and indicates the background. Capital letters A to G correspond to the main structures and are explained in the text. Dotted lines indicate the magnitude of the PCI effect. The vertical bars in the lower part of the panel indicate the energies of the Ca$_{aq}^{3+}$(1s$^{-1}$) $\rightarrow$ Ca$_{aq}^{4+}$(2p$^{-2}$($^{1}$S, $^{1}$D, and $^3$P)) Auger transitions without PCI shift. The energy position of the  Ca$_{aq}^{3+}$(1s$^{-1}$) $\rightarrow$ Ca$_{aq}^{4+}$(2p$^{-2}$($^1$S)) transition is obtained by an extrapolation as shown in Fig. \ref{fig:fg3}b}
    \label{fig:fg1}
\end{figure}

The X-ray absorption spectrum of hydrated calcium shown in the right panel of Fig.\ \ref{fig:fg1} is 
identical to that presented in Ref. \citep{saisopa2020:146984}. Based on our calculations, the first structure at h$\nu$ = 4040 eV is assigned to the Ca$_{aq}^{2+}$(1s$^{-1}$3d) resonances, in full agreement with Ref. \citep{saisopa2020:146984}. Note that here and in the following the character of the excited-state orbitals of the micro-solvated clusters is derived from their overlap with the atomic orbitals of isolated Ca$_{aq}^{2+}$ ions, and thus, the core-excited states below threshold are labeled using the dominant virtual orbitals of the metal ion in the excited-state wave function. For the model geometry considered in this work, the states below threshold are rather atomic-like in nature. However, the states above threshold are not pure atomic states, but they rather possess a non-negligible contribution from water orbitals (see Fig. \ref{fig:si1}) and we will therefore use the notation charge-transfer-to-solvent (CTTS) states similar to Ref. \citep{muchova2024:8903}.

In detail, the 3d orbital matches well those of the Ca$_{aq}^{2+}$ ion and is strongly localized at the ionic site, see Fig.\ \ref{fig:si2} of the supplementary material. The excitations to the Ca$_{aq}^{2+}$(1s$^{-1}$3d) states are dipole-forbidden in the highly symmetric [Ca(H$_{2}$O)$_6$]$^{2+}$ (octahedral) and [Ca(H$_{2}$O)$_8$]$^{2+}$ (cubic) micro-solvated clusters, which both show inversion symmetry at the Ca site. 
However, the transition can gain intensity in two different ways. First, due to the high excitation energy quadrupole contributions cannot be excluded\citep{arrio2000:454}. Second, lower symmetry arrangements without inversion symmetry at the Ca site, like the [Ca(H$_{2}$O)$_7$]$^{2+}$ cluster, as well as in [Ca(H$_{2}$O)$_6$]$^{2+}$ and [Ca(H$_{2}$O)$_8$]$^{2+}$ clusters with distorted symmetry of the water molecules lead to a small admixture of p-character to the 3d orbital and render the transition dipole-allowed.

The analysis of Ca$^{2+}$ K-pre-edge XAS recorded for different compounds reveals that in centrosymmetric environments, such as six-coordinate complexes, dipole-forbidden 1s$^{-1}$3d excitations dominate as quadrupole-allowed transitions\citep{martind2015:1283}. In contrast, symmetry breaking in seven- or non-cubic eight-coordinate complexes introduces p-d orbital mixing, enabling dipole-allowed transitions (1s$^{-1}$(3d + np)) and increasing pre-edge intensity by about one order of magnitude due to the combined quadrupole and dipole contributions. 
Both dipole- and quadrupole-allowed transitions can contribute to the pre-edge structure in this case, with dipole-allowed transitions likely being more prominent due to the asymmetrical arrangement of water molecules surrounding the metal ion. The average coordination number of solvated Ca$^2_{aq}$ is between 7 and 8 \citep{jalilehvand2001:431,megyes04:7261,tongraar10:10876}.
These transitions are located directly below the Ca$_{aq}^{2+}$ 1s$^{-1}$ ionization threshold at 4043.6 eV as obtained by X-ray photoelectron spectroscopy (XPS), see Fig.\ \ref{fig:xps}.

A comparison of the X-ray absorption spectra of K$_{aq}^{+}$ and Ca$_{aq}^{2+}$ below the 
1s$^{-1}$ ionization threshold reveals significant differences. In the Ca$_{aq}^{2+}$ spectrum, only one excitation is observed, namely the Ca$_{aq}^{2+}$ (1s$^{-1}$3d) resonance, located about 3.6 eV below the threshold. All Ca$_{aq}^{2+}$ (1s$^{-1}$np) resonances are embedded in the continuum. In contrast, the order of these states in K$_{aq}^+$  is reversed. The lowest resonance is K$_{aq}^{+}$ (1s$^{-1}$4p), followed by K$_{aq}^{+}$ (1s$^{-1}$3d), both found within 1 eV below the K 1s threshold \citep{miteva18:4457}.

This difference in the ordering of core-excited states between K$_{aq}^{+}$ and Ca$_{aq}^{2+}$ can be explained by examining the bare ions. In the metal ion sequence K$^{+}$, Ca$^{2+}$, and Sc$^{3+}$, the 1s$^{-1}$3d state is always lower in energy than the 1s$^{-1}$4p state, with the energy gap increasing with the nuclear charge. Using the Z+1 approximation and data from NIST \citep{NIST_ASD}, the splitting is 1.4 eV for K$^{+}$, 7.7 eV for Ca$^{2+}$, and 15.9 eV for Sc$^{3+}$.

Since the 3d orbital is highly localized around the ion, the energy of the 1s$^{-1}$3d state is only weakly affected by the surrounding water molecules (see Fig. \ref{fig:si1} and Fig. 4 in Ref. \citep{bjorneholm92:1892}). In contrast, the larger 4p orbital interacts more strongly with the water orbitals of the first solvation shell (see Fig. \ref{fig:si2}). Consequently, the energy of the 4p core-excited states in the solvated ion drops by 2.8 eV for Ca$^{2+}$ and by 1.6 eV for K$^{+}$ compared to the bare ion, with this shift being more significant for the 4p states than for the 3d states. In K$^{+}$, unlike in Ca$^{2+}$, these shifts result in an inversion of the order of the core-excited states.

The X-ray absorption spectrum is dominated by the main peak at 4050.8 eV and a shoulder at 4046.5 eV, which are 10.8 eV and 6.5 eV above the transitions to the Ca$_{aq}^{2+}$(1s$^{-1}$3d) states, respectively. Independent of the number of water molecules $m$, the calculations show three peaks, which are located approximately 5 eV, 8 eV, and 9.5 eV above the transitions to the Ca$_{aq}^{2+}$(1s$^{-1}$3d) states, with the peak at about 8 eV being the dominant one; for more details, see the SI. We relate the theoretically obtained peaks at 8 eV and 9.5 eV above the transitions to the Ca$_{aq}^{2+}$(1s$^{-1}$3d) states to the main peak at 10.8 eV, i.e.\,they are not resolved in the experimental data. Moreover, the shoulder in the experimental spectrum is assigned to the theoretical peak at 5 eV above the transitions to the Ca$_{aq}^{2+}$(1s$^{-1}$3d) states. Based on this assignment and the theoretical results, the shoulder in the experimental spectrum at 4046.5 eV is due to transitions to CTTS core excited states, with a Ca np (n=4, 5) character, see theoretical results in the SI. The radial dimension of the natural orbitals occupied by the excited electron in question is larger than the Ca – O distance, i.e.\,they extend well into the surrounding water. The main peak at 4050.8 eV as well as its high-energy part can be assigned to the next two sets of core-excited CTTS states (between 4048 and 4051 eV on Fig. \ref{fig:si1}). These states are of np-metal character (n$ = 4, 5, 6$) with a large contribution of solvent molecular orbitals (see  Fig. \ref{fig:si1}). 

At 4070.4 eV, a small structure is observed, as noted in Refs.\ \citep{dangelo04:11857,fulton2003:4688,fulton2006:094507}   and it is attributed to KM$_{2,3}$ double excitations. However, no precise assignment has been made. In the simpler case of isoelectronic Ar in the gas phase, double excitations are also detected approximately 20 eV above the 1s threshold. For this atom, different calculations reveal strong electron-correlation effects and reproduce the spectrum reasonably well \citep{kavcic2009:143001,sukhorukov87:1677,zitnik2023:053113}, though they differ in the contributions of various configurations. The most significant contributions are the configurations  1s$^{-1}$3p$^{-1}$4p$^2$, 1s$^{-1}$3p$^{-1}$4s4d, and 1s$^{-1}$3p$^{-1}$3d$^2$. Therefore, we can expect similar contributions in the present case of Ca$_{aq}^{2+}$. 

\subsection{The Ca$_{aq}^{2+}$ KL$_{2,3}$L$_{2,3}$ Auger spectra}

We recorded resonant and non-resonant KL$_{2,3}$L$_{2,3}$  Auger spectra at selected energies both in the vicinity of the Ca$_{aq}^{3+}$ (1s$^{-1}$) ionization threshold, as well as up to 1 keV above. Selected spectra are shown in the left panel of Fig. \ref{fig:fg1}, with the intensity levels on the right side of the panel indicating the photon energies used. The main structures in the spectrum recorded at h$\nu$ = 5000 eV are labeled with the letters A to E. The peak shifts indicated for the spectra recorded at lower photon energies are due to the PCI effect, which will be discussed in detail further below. Letter F indicates the position of the resonant Auger transition for excitations just below threshold, and the letter G indicates the resonant Auger decay subsequent to KM$_{2,3}$ double excitations at 4070.4 eV.

\subsubsection{Resonant Auger spectra below threshold}

\begin{figure}
    \centering
    \includegraphics[width=0.5\linewidth]{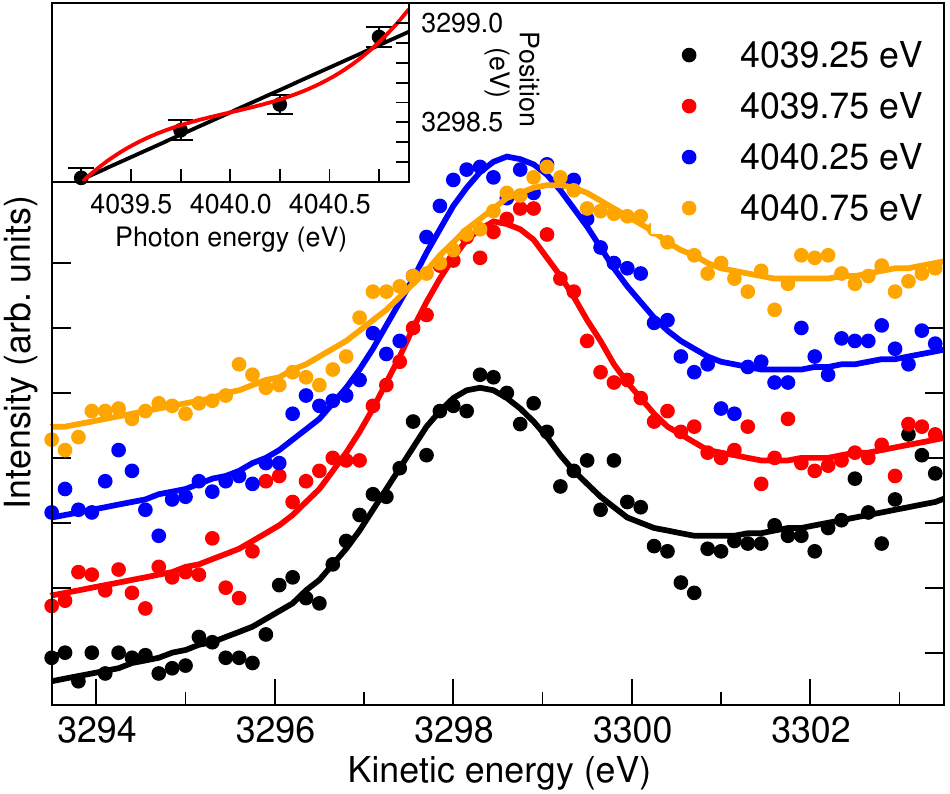}
    \caption{The Ca$_{aq}^{2+}$(1s$^{-1}$3d) $\rightarrow$ Ca$_{aq}^{3+}$(2p$^{-2}$($^1$D)3d) resonant Auger spectra recorded at photon energies between 4039.25 to 4040.75 eV. The solid lines through the data points represent the fit results. The inset displays the dispersion of the peak position with photon energy. The black line in the inset is a fit to a linear dispersion with a slope of 0.45 and the red spline indicates an S-shape behavior as predicted by theory, for details see text.}
    \label{fig:fg2}
\end{figure}

Since the 1s$^{-1}$3d excited state is located below threshold it is expected to undergo resonant Auger decay producing Auger electrons of kinetic energies around 3298 eV (see feature F in the left panel of Fig. \ref{fig:fg1}). The resulting feature F is assigned to the Ca$_{aq}^{2+}$(1s$^{-1}$3d) 
$\rightarrow$ Ca$_{aq}^{3+}$(2p$^{-2}$($^1$D)3d) transition and indicates that the excited electron remains at the Ca$^{2+}$ site during the core-hole lifetime. The four spectra recorded at excitation energies between 4039.25 and 4040.75 eV, see Fig.\,\ref{fig:fg2}, are fitted with a Lorentzian peak of FWHM 0.81 eV \citep{krause79:329} to simulate the Ca$_{aq}^{3+}$(1s$^{-1}$) core-hole lifetime, which is convoluted with a Gaussian to simulate the experimental resolution and the inhomogeneous broadening caused by the different and variable surroundings of the calcium ion; the Gaussian width was a free fit parameter.   

Surprisingly, the dispersive behavior of the resonant Auger structures is not linear as shown in the inset of Fig.\ \ref{fig:fg2}. Generally, this behavior depends on the ratio between photon bandwidth and natural lifetime\citep{gelmukhanov96:3960}. 
With very high resolution, i.e., when the photon bandwidth is much smaller than the lifetime broadening, the dispersion should have a slope of 1. In contrast, with very low photon resolution, the slope should approach zero at the resonance energy.

In the present case, the black line in the inset of Fig. \ref{fig:fg2} indicates a slope of approximately 0.45. This can be explained by the photon bandwidth of 0.6 eV being comparable to the lifetime width of 0.81 eV. Regardless of the photon bandwidth, far from the resonance position, the slope must reach 1 due to energy conservation. When the photon bandwidth is not much smaller than the lifetime broadening, this results in an overall S-shaped dispersion, as shown by the red cubic spline. The exact behavior is detailed by Gel’mukhanov and \AA{}gren\citep{gelmukhanov96:3960}.  Experimentally, this effect has been observed by Kukk et al. \citep{kukk97:1481} for the Kr (3d$^{-1}$5p) $\rightarrow$ Kr$^{+}$(4p$^{-2}$($^1$D)5p) resonant Auger decay. A more systematic study of this effect for Ca$_{aq}^{2+}$ would require additional data and is beyond the scope of this work. 

As mentioned earlier, the Gaussian width of the peaks includes the experimental resolution and inhomogeneous broadening due to variations in the calcium ion’s environment. From our results, we estimate an inhomogeneous broadening of 1.8(2) eV, which reflects the variation in Auger energies caused by different surroundings.

\subsubsection{The normal Auger spectra above threshold}

\begin{figure}
    \centering
    \includegraphics[width=0.5\linewidth]{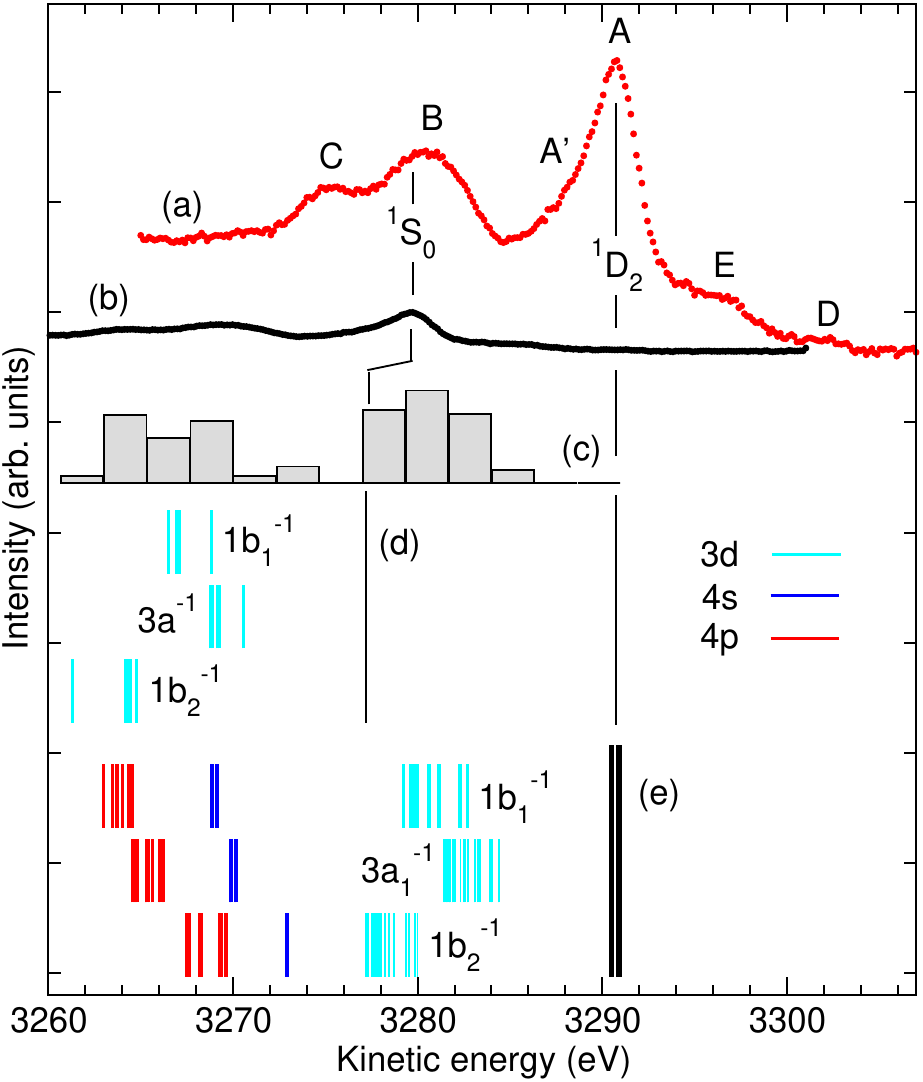}
    \caption{Experimental and theoretical results for the KL$_{2,3}$L$_{2,3}$ Auger decay of Ca$_{aq}^{3+}$(1s$^{-1}$). (a) Experimental spectrum recorded at h$\nu$ = 5 keV. The curve in (b) estimates the contributions of the Auger decay to the configuration 2p$^{-2}$($^{1}$S) by using the data presented in (a), however, shifted by 11.1 eV to lower values and divided in intensity by 7.4; for more details, see text. Panel (c) shows a histogram of the density of transitions from the 1s$^{-1}$ level to the CT states. Here all transitions leading to the 2p$^{-2}$($^{1}$D) double core hole are weighted with one and those leading to the 2p$^{-2}$($^{1}$S) double core hole state with 1/7.4 $\approx$ 0.135 in order to take the lower intensities of the latter transitions into account. The colored vertical lines in (d) and (e) represent the theoretical kinetic energies for the Auger decays to the charge transfer states Ca$_{aq}^{3+}$(2p$^{-2}$($^{1}$S)nl)H$_2$O$^{+}$(val$^{-1}$) and Ca$_{aq}^{3+}$(2p$^{-2}$($^{1}$D)nl)H$_2$O$^+$(val$^{-1}$), respectively. The different panels in (d) and (e) indicate the hole in the water molecule for the CT state. The long black vertical lines indicate the theoretical energy positions of the Auger decays to the 2p$^{-2}$($^{1}$S) and 2p$^{-2}$($^{1}$D) final states. The energy positions for the theoretical data are shifted so that the experimental and theoretical energy positions for the Ca$_{aq}^{3+}$(1s$^{-1}$) $\rightarrow$ Ca$_{aq}^{4+}$(2p$^{-2}$($^{1}$D)) transitions match.}
    \label{fig:fg3}
\end{figure}

In contrast to the Auger spectra recorded below the Ca$_{aq}^{3+}$(1s$^{-1}$)  ionization threshold, the spectra above threshold exhibit signatures of complex photon-energy-dependent Auger processes that involve both local and environmental effects. We first examine the Auger spectrum recorded at 5 keV photon energy, approximately 1 keV above threshold (red curve at the top of Fig.\ \ref{fig:fg1}), where the influence of post-collision interaction (PCI) is negligible. The most intense peak, labeled A, at 3290.8 eV kinetic energy, is attributed to the normal Auger transition Ca$_{aq}^{3+}$(1s$^{-1}$) 
$\rightarrow$ Ca$_{aq}^{4+}$(2p$^{-2}$($^1$D)), as indicated in the lower part of the figure. On the low kinetic-energy side, an asymmetric feature labeled A’ is observed. This asymmetry in the main line could arise from several factors, including the onset of an asymmetric background in this kinetic energy range (e.g., the Auger spectrum of K$^{+}$ in water reported in Ref. \citep{ceolin17:263003}), Auger decays to final states involving charge transfer from neighboring water molecules—which, in a realistic water environment, could occur at higher energies than those predicted by our model calculations on microsolvated Ca$^{2+}$ clusters with one or two water molecules (see Fig.\ \ref{fig:fg3})—or contributions from the Auger decay of Ca$^{3+}$ (1s$^{-1}$) satellite states observed in the XPS spectrum (see Fig.\ \ref{fig:xps}), as indicated by our computed Auger spectrum (see Fig.\ \ref{fig:si_auger_theory}).

At the high kinetic-energy side of the A peak two less intense structures are visible, namely structure D at 3302 eV, which is attributed to the Ca$_{aq}^{3+}$(1s$^{-1}$) $\rightarrow$ Ca$_{aq}^{4+}$(2p$^{-2}$($^3$P)) transition, and structure E at 3296.5 eV whose origin is different from that of the other peaks and will be discussed further below. The weak transition D is not distinguishable from the background for h$\nu$ = 4044 eV and 4046 eV but clearly visible at 3302 eV kinetic energy using h$\nu$ = 5000 eV.
The second and third most intense structures are observed at 3280.4 eV and 3275.5 eV kinetic energies, and are labeled as B and C, respectively. They result from the superposition of several components. One of them is the Auger decay to the Ca$_{aq}^{4+}$(2p$^{-2}$($^{1}$S)) final state. Since the energy position of the $^{1}$S component of the 2p$^{-2}$ configuration cannot be precisely determined in the experiment, we estimated the energy separation of the $^1$D and $^{1}$S components to 11.1 eV (for details, see SI). Note that the KL$_{2,3}$L$_{2,3}$ normal Auger transitions of the hydrated calcium dication are pure inner-shell processes and, therefore, hardly influenced by the chemical surrounding. The $^{1}$S$_0$/$^1$D$_2$ Auger intensity ratio for argon is about 1:7.4 (Ref. \citep{asplund77:268}), and we expect a similar value for hydrated calcium. Using this intensity and the above discussed splitting, we simulate, based on the Ca$_{aq}^{3+}$(1s$^{-1}$) $\rightarrow$ Ca$_{aq}^{4+}$(2p$^{-2}$($^1$D)) normal Auger transition, the spectral contributions of the Ca$_{aq}^{3+}$(1s$^{-1}$) $\rightarrow$ Ca$_{aq}^{4+}$(2p$^{-2}$($^{1}$S)) transition, see red and black curve in Fig.\ \ref{fig:fg3}. The result clearly shows that the contributions of the $^{1}$S component estimated by the black curve can only explain a fraction of the intensity observed in the kinetic energy range where the Auger transition to the $^{1}$S final state is expected. 

As an additional contribution to the Auger spectra in the kinetic energy range below 3284 eV -- approximately 6 eV below peak A -- we anticipate the presence of Auger electrons that have undergone energy loss processes due to interactions with water molecules, similar to the behavior observed in the photoelectron spectra. For the present complex Auger spectrum with a multitude of different transitions we expect these contributions to be minor and relatively unstructured, forming part of the background. However, they may become more significant in Auger spectra of ions in liquids that exhibit fewer distinct Auger peaks.

It is interesting to note that the present KL$_{2,3}$L$_{2,3}$ Auger spectrum for Ca$_{aq}^{2+}$ matches well the KL$_{2,3}$L$_{2,3}$ Auger spectra for solid CaO, CaCO$_3$, and CaSO$_4$ \citep{nishikida78:49}. In Ref. \citep{nishikida78:49}, all peaks not identified as Ca$_{aq}^{3+}$(1s$^{-1}$) $\rightarrow$ Ca$_{aq}^{4+}$(2p$^{-2}$) transitions are assigned to charge transfer states, where an electron from the oxygen migrates to the Ca ion. Our further analysis supported by theoretical calculations indeed shows the presence of ultrafast charge transfer processes between the solvent and the metal ion.

To gain deeper insight into the nature of charge transfer (CT) processes following KL$_{2,3}$L$_{2,3}$ Auger decay in an aqueous solution of Ca$^{2+}$, we computed the energy positions of the Auger final states for the mono-hydrated Ca-ion, Ca$^{3+}$(H$_{2}$O). Panels (d) and (e) of Fig.\ \ref{fig:fg3} show the calculated kinetic energies for Auger transitions to the charge transfer states: Ca$_{aq}^{3+}$(2p$^{-2}$($^{1}$S)nl)H$_{2}$O$^+$(val$^{-1}$) and Ca$_{aq}^{3+}$(2p$^{-2}$($^1$D)nl)H$_{2}$O$^+$(val$^{-1}$), respectively.

For comparison with experimental data, the computed binding energies were shifted to align with the energy of the experimental $^1$D main line. The calculated energy splitting between the ¹D and $^{1}$S main lines is approximately 13.5 eV, which is in good agreement with the extrapolated experimental value of 11.1 eV. The contributions from transitions to the ³P final states are expected to be minimal and are therefore not shown.

The vertical lines in panels (d) and (e) are color-coded to represent the main character of the metal orbital occupied during the charge transfer process. Since these calculations do not provide Auger intensities, in Fig. \ref{fig:fg3}(c) we present a histogram of Auger transitions to CT states for comparison with experiment. To plot the histogram, the intensities of the final states with $^{1}$D symmetry were set to 1, while those with $^{1}$S symmetry were reduced to $1/7.4 \cong 0.135$ to reflect their generally lower intensity. The Ca$_{aq}^{3+}$(1s$^{-1}$) $\rightarrow$ Ca$_{aq}^{4+}$(2p$^{-2}$) transitions are excluded from this histogram. The histogram aligns well with the experimental spectrum’s intensity, enabling the assignment of specific spectral features. Based on this histogram, we attribute structure B to transitions to the Ca$_{aq}^{3+}$(2p$^{-2}$($^1$D)3d)H$_{2}$O$^+$(val$^{-1}$) CT states as well as to a minor extent to the Ca$_{aq}^{3+}$(2p$^{-2}$($^{1}$S)) state, and peak C primarily to transitions to the Ca$_{aq}^{3+}$(2p$^{-2}$($^1$D)4p/4s)H$_{2}$O$^+$(val$^{-1}$) CT states. 

A summary of the interpretation of the KL$_{2,3}$L$_{2,3}$ Auger spectrum recorded at h$\nu$ = 5 keV is shown in Table \ref{tab:tab1}. In short, we explained the structures B and C as transitions to 2p$^{-2}$($^1$D) charge-transfer states by using an atomic orbital picture. All these states are associated with a charge transfer from the water valence orbitals to the 3d, 4s, and 4p virtual orbitals of the core-ionized calcium di-cation. The structures B and C also possess minor contributions of final states, where the 2p$^{-2}$ configuration is coupled to the $^{1}$S. These charge transfers occur during the Auger process, which provides the energy for the ionization of the water molecule, i.e. the energy is balanced by the Auger electron.

\begin{table*}
\centering \caption{A summary of the features A to F observed in the Auger spectrum of Ca$^{2+}_{aq}$, their energy and the respective transition assignment in this work.}
\footnotesize
\begin{tabular}{llll} \\ \hline \hline
Peak  & Kinetic   & Corresponding Auger transitions & Decay processes \\ 
notation & energy  &  &  \\ \hline
A & 3290.8 & Ca$^{3+}_{aq}$(1s$^{-1}$) $\rightarrow$ Ca$^{4+}_{aq} $(2p$^{-2}$($^1$D)) & Normal Auger (NA) to 2p$^{-2}$($^1$D) \\ \\

A’ & 3287.5 & Ca$^{3+}_{aq}$(1s$^{-1}$) $\rightarrow$ Ca$^{3+}_{aq}$(2p$^{-2}$($^1$D)3d) H$_2$O$^+$ & NA to 2p$^{-2}$($^1$D) + CT 
from H$_2$O \\

     &  &  &  Asymmetric background \\

     &  & Ca$^{2+}_{aq}$ (1s$^{-1}$nl) H$_2$O$^+$ $\rightarrow$ Ca$^{3+}_{aq}$(2p$^{-2}$n'l') H$_2$O$^+$ &   Auger decay of photoelectron satellites \\ \\

B & 3280.5 & Ca$^{3+}_{aq}$(1s$^{-1}$) $\rightarrow$ Ca$^{3+}_{aq}$(2p$^{-2}$($^1$D)3d) H$_2$O$^+$ & NA to 2p$^{-2}$($^1$D) + CT from H$_2$O \\

 &  & Ca$^{3+}_{aq}$(1s$^{-1}$) $\rightarrow$ Ca$^{4+}_{aq} $(2p$^{-2}$($^1$S)) & NA to 2p$^{-2}$($^1$S) \\ \\

C & 3275.0 & Ca$^{3+}_{aq}$(1s$^{-1}$) $\rightarrow$ Ca$^{3+}_{aq}$(2p$^{-2}$($^1$D)4s/4p) H$_2$O$^+$ & NA to 2p$^{-2}$($^1$D) + CT from H$_2$O \\ \\

D & 3302.0 & Ca$^{3+}_{aq}$(1s$^{-1}$) $\rightarrow$ Ca$^{4+}_{aq}$(2p$^{-2}$ ($^3$P)) & NA to 2p$^{-2}$($^3$P) \\ \\

E & 3296.5 & Ca$^{2+}_{aq}$(1s$^{-1}$4s) $\rightarrow$ Ca$^{3+}_{aq}$(2p$^{-2}$ 4s)  & Electron capture + Auger decay \\

 & & Ca$^{3+}_{aq}$(1s$^{-1}$) Cl$^{-}_{aq}$ $\rightarrow$ Ca$^{3+}_{aq}$(2p$^{-2}$ 3d) Cl$^{0}_{aq}$ & NA + CT from Cl$^- $ \\
 
 &  & Ca$^{2+}_{aq}$ (1s$^{-1}$nl) H$_2$O$^+$ $\rightarrow$ Ca$^{3+}_{aq}$(2p$^{-2}$n'l') H$_2$O$^+$ & Auger decay of photoelectron satellites \\ \\
 
F & 3298.7 & Ca$^{2+}_{aq}$(1s$^{-1}$3d) $\rightarrow$ Ca$^{3+}_{aq}$(2p$^{-2}$($^1$D)3d) & Resonant Auger \\ \hline\hline
\end{tabular}
\label{tab:tab1}
\end{table*}
 
As discussed above, notable differences are observed in the X-ray absorption spectra of K$^{+}_{aq}$ and Ca$^{2+}_{aq}$. Similarly, substantial differences are evident between the KL$_{2,3}$L$_{2,3}$ normal Auger spectra of the two ions. Since the 2p$^{-2}$ final states of the metal ions as well as the CT states result from the decay of the 1s$^{-1}$ core-ionized state, the Auger electron energies of the main lines and CT peaks indicate the energy ordering of the final states. In the case of K$^{+}$, the $^1$D and $^{1}$S  main lines are clearly visible in the Auger spectrum above threshold, and transitions to CT states of the type 
K$_{aq}^{2+}$(2p$^{-2}$($^1$D)3d)H$_{2}$O$^+$(val$^{-1}$) appear as a shoulder 
of the $^1$D main line. CT transitions to the higher-lying 4s and 4p virtual orbitals of the metal ion are not observed in the Auger spectrum above threshold. In contrast, due to the higher charge of Ca$^{2+}$, the transitions to CT states of the type Ca$_{aq}^{3+}$(2p$^{-2}$ 3d/4s/4p)H$_{2}$O$^+$  are observed in the vicinity of those to the Ca$^{2+}_{aq}$(2p$^{-2}$) $^{1}$D and $^{1}$S states. Since these transitions are located in the same energy range as the main Auger transitions, and due to the efficiency of the CT process, the $^{1}$S main Auger line in the normal Auger spectrum of Ca$_{aq}^{2+}$ is obscured by the Auger transitions to these CT states.

\subsubsection{Photon-energy dependent processes: PCI, double excitations, and charge transfer}

The Auger spectrum recorded at 5 keV presents signatures of normal Auger and charge transfer processes. However, a single spectrum cannot capture the photon-energy dependent processes. Such effects can be identified by considering the spectra recorded at a range of intermediate photon energies between 4042 eV and 4140 eV. The spectra taken between 4042 eV and 4052 eV illustrate the evolution of the KL$_{2,3}$L$_{2,3}$ Auger decay while crossing the Ca$_{aq}^{2+}$ 1s ionization threshold at h$\nu$ = 4043.6 eV.

A notable photon-energy dependent effect is a clear shift of all structures towards higher kinetic energies as compared to the spectrum recorded at 5 keV. In the left panel of Fig.\ \ref{fig:fg1}, this shift is indicated by dotted lines for structures A to D, and it is attributed to the PCI effect (discussed in more detail in the SI). In contrast to isolated atoms and small molecules, where the PCI shift decreases continuously and rapidly with the kinetic energy of the photoelectron, the structures in the present case exhibit an almost constant PCI shift of about 1.2 eV within the energy range of 4042 eV to 4052 eV, relative to the spectrum measured at a photon energy of 5 keV. Such a behavior was recently detected in the organic polymers polythiophene and poly(3-hexylthiophene-2,5-diyl), commonly known 
as P3HT\citep{velsquez2023:013048}. The spectra measured at h$\nu$ = 4070.4 eV and h$\nu$ = 4140 eV exhibit a smaller PCI shift. This effect smoothly disappears when the photoelectron energy increases, with the PCI shift at h$\nu$ = 4070.4 eV amounting to about 0.5 eV relative to the spectrum taken at 5 keV. Moreover, in addition to the induced shift, this effect causes an asymmetry in the shape of the photoemission and Auger peaks. For the photoemission peak this asymmetry is visible on the low kinetic-energy side, whereas for the Auger peaks, it is visible on its high kinetic-energy side.

At h$\nu$ = 4070.4 eV, see right panel of Fig.\ \ref{fig:fg1}, two overlapping processes are identified, namely the Ca$_{aq}^{3+}$ (1s$^{-1}$)  direct ionization and KM$_{2,3}$ double excitations. In addition to the normal Auger decay and CT processes already discussed above, we observe a weak structure labeled G, which is partially overlapping with peak E but only present at this photon energy. We therefore assign G to resonant Auger decays of doubly excited states.

The Auger spectra recorded at photon energies directly above the ionization threshold show that transition E 
exhibits photon-energy dependent intensity variations different from transitions A – D and A’. This structure
is absent in the spectrum recorded at h$\nu$ = 4044 eV, i.e.\,0.4 eV above the Ca$_{aq}^{2+}$ 1s ionization 
threshold. It appears at photon energies 4-5 eV higher and is also present far above threshold at 5 keV. This distinct photon-energy dependence, unlike that of structures A to D clearly shows that feature E must be of different origin. In the following, we discuss four possible explanations for feature E ruling out the first. Based on our current experimental and theoretical data, we cannot fully confirm or exclude any of the remaining three hypotheses.

First, it is interesting to note that the photon energy onset of peak E coincides with the CTTS states of 1s$^{-1}$np character above threshold. This suggests that it might be related to the decay of such states. A resonant Auger process is not confirmed by the experimental data due to the lack of dispersive behavior of feature E and because it does not fade away as the photon energy increases. In the following we give reasons against this explanation. For this we have to distinguish between Auger final states with one continuum electron (namely resonant Auger or electron recapture) and two continuum electrons (normal Auger including PCI). In principle, final states with only one continuum electron, such as those populated via a recapture of the excited continuum electron during the Auger decay into a bound 2p$^{-2}$np state, are possible. However, due to simple energy-conservation arguments such recapture processes should lead to spectral features which obey h$\nu$ - E$_{kin}$ $<$ I.P. - E$_{Auger}$($^1$D) = 745.8 eV, where E$_{kin}$ is the kinetic energy of the Auger electron, and E$_{Auger}$($^1$D) is the position of the $^1$D main line without PCI shift. This inequality does not hold for peak E. As a consequence, for the photon and kinetic energies of peak E, the photoelectron and the Auger electron are both continuum electrons so that their interaction should be fully described by the PCI effect discussed above. Because of this we consider it highly unlikely that peak E is related to the decay of the CTTS states.

A second possibility for the formation of structure E is the Auger decay of photoelectron satellites of the type Ca$_{aq}^{2+}$(1s$^{-1}$nl)H$_{2}$O$^{+}$. This possibility is supported by the theoretical Auger spectrum computed by taking into account the photoelectron satellites (see Fig. \ref{fig:si_auger_theory}). As can be seen from the theoretical Ca 1s photoelectron spectrum on Fig.\ \ref{fig:xps}, such satellite states are found only at higher energies, i.e.\ at least 10 eV above the 1s ionization threshold. As previously discussed, the lower-energy part of the XPS spectrum starting at 4050 eV is dominated by electron-energy loss due to IPES leading excitation of water molecules from the medium. Therefore, the decay of photoelectron satellites can only contribute to peak E at photon energies above 4054 eV and cannot explain the structure between its onset and this value.

An unscattered photoelectron with a kinetic energy of about 4 eV, i.e.\ at the onset of peak E, travels within the core-hole lifetime of 0.8 fs by about 10 \AA . From this we conclude that peak E at photon energies between the onset at 4048 eV and the possible Auger decays of photoelectron satellites at 4054 eV must be due to processes, which are prevented by a photoelectron close to the Ca site, namely as long as the surrounding sees a Ca$^{2+}$ ion. Instead, peak E has to be due to the fact that the surrounding sees a Ca$^{3+}$  ion.

A possible explanation is that feature E originates from a CT process from the counterion Cl$_{aq}^{-}$ to the metal ion which in a simple model is estimated to be found about 2 eV above the kinetic energy of peak A (see Fig.\ \ref{fig:si4}). The charge-transfer occurs on the timescale of the Auger process once the photoelectron is sufficiently far away from the Ca site, and leads to the population of Ca$_{aq}^{3+}$(2p$^{-2}$ 3d)Cl$_{aq}^{0}$ states. While no evidence of contact ion pairs has been observed in a 1M aqueous solution of CaCl$_2$ \citep{megyes04:7261,todorova2008:779}, solvent-shared and double-solvent-shared ion pairs have been found in a 1.8M solution\citep{friesen2019:891}. The separation between Ca and Cl in the 1M solution of CaCl$_2$ studied in this work is approximately 3.5-5.1\,\AA~\citep{todorova2008:779}. CT in this case should take place over this distance, and is favored by the Coulomb attraction between Ca$^{3+}$ and Cl$^{-}$ ions. 

A non-negligible overlap of Cl$^-$ and Ca$_{aq}^{3+}$(2p$^{-2}$ 3d) can be expected since the size of the 3d orbital in Ca$^{2+}$(1s$^{-1}$ 3d) is larger than 2 \AA~(see Fig.\ \ref{fig:si2}), and
the ionic radius of Cl$^-$ in water is about 1.8 \AA\citep{markus88:1475}. Therefore, the sum of the two is larger than the
minimal Ca-Cl distance of 3.5 \AA, so that even the higher charge of the metal ion and, thus, smaller 3d radius in Ca$_{aq}^{3+}$(2p$^{-2}$ 3d) should not fully exclude overlap.

The photon energy dependence of this CT process results from the screening of K-shell ionized Ca$^{3+}$ by the low-kinetic-energy photoelectron (KE $\le$ 4 eV). At a KE of 4 eV an unscattered  photoelectron travels about 10 \AA~within the core-hole lifetime of 0.81 fs. Therefore, the electron is sufficiently far from the hydrated ion pair, and it no longer screens the metal ion thus allowing for the charge-transfer to occur. Elastic scattering of an electron in water was found to be the predominant mechanism for low-KE electrons\citep{sinha2021:5479}. The elastic mean free path of an electron of a KE of 10 eV was found to be 0.56 ($\pm$ 0.01) nm\citep{schild2020:1128}, which is on the order of magnitude of the Ca$-$Cl distance.

Another possibility for the emission of high-KE Auger electrons is the attachment of a free electron from the surrounding medium to the metal ion during the core-hole lifetime.
This occurs when the metal ion, no longer screened by the photoelectron due to its sufficiently large separation, see above, attracts the free electron. In this case, the free electron is captured in a low-n Ca$^{2+}_{aq}$(1s$^{-1}$nl) state while the attachment energy is balanced by the emission of a photon\citep{jahnke2020:11295} or a direct energy transfer to the water like the excitation of vibrational modes. In this context, we assume that the electron is captured in the lowest core excited state with s-character, namely the Ca$^{2+}_{aq}$(1s$^{-1}$4s)  excitation since its energy position is, according to the present calculations, very close  to the ionization threshold. We want to point out that the calculated size of about 6 \AA \ for the excited electron orbital with a predominant 4s metal character in a [Ca(H$_{2}$O)$_7$]$^{2+}$ cluster (Fig.\ \ref{fig:si2}) is smaller than the distance of 10\,\AA~covered by an unscattered photoelectron with a kinetic energy of 4 eV within the core-hole lifetime of 0.81\,fs which represents the onset of peak E and supports the suggested process for its formation.


In Ref. \citep{nishikida78:49} the authors assign the satellite states of solid CaO, CaCO$_3$, and CaSO$_4$ equivalent to features B, C and E in this work, to charge transfer states populated in the Auger decay - where a 4s, 4p and 3d orbital of Ca is involved in the charge transfer process. The authors generally assume that they are shake-up satellites following the Auger decay of the Ca$^{3+}$(1s$^{-1}$) state. Our calculations demonstrate that features B and C indeed correspond to charge-transfer states populated in the Auger decay of the Ca$^{3+}$(1s$^{-1}$) state. However, feature E shows a different photon-energy dependence and we suggest three possible processes for its formation. Based on our calculation, Auger decays of photoelectron satellites can explain this feature above 4054 eV. Below this value of the photon energy, a CT from a Cl$^{-}$ ion in the second hydration shell and an electron capture into an unoccupied orbital of Ca are possible processes leading to the formation of feature E. They can occur once the photoelectron has moved away from the ionization site. The results show that accounting for the photon-energy dependence of the Auger spectra is essential for the correct identification of the underlying ultrafast processes.


\section{Summary and conclusions}\label{sec:concl}
In conclusion, we investigated the range of ultrafast electronic decay processes that occur in aqueous CaCl$_2$ solutions following Ca 1s$^{-1}$ core-level ionization using X-ray absorption, photoelectron, and Auger spectroscopy. These experimental results were further supported by high-level quantum-chemical calculations.
Below the Ca 1s$^{-1}$ ionization threshold, a single resonance was observed, which is attributed to the population of the Ca$_{aq}^{2+}$(1s$^{-1}$3d) core-excited states. The resonant Auger decay of these states results in a distinct feature in the Auger spectrum. Surprisingly, the resonant Auger process does not exhibit the typical linear dispersion with photon energy but instead shows a slope of approximately 0.45 relative to the photon energy. This effect is explained with similar values for the experimental resolution and the lifetime width as known from studies of dilute matter.

Above the Ca 1s$^{-1}$ ionization threshold, the conventional Auger decay  Ca$_{aq}^{3+}$(1s$^{-1}$) $\rightarrow$ Ca$_{aq}^{4+}$(2p$^{-2}$($^{1}$D,$^{1}$S)) is accompanied by a series of charge transfer (CT) processes, where an electron is transferred from the surrounding water molecules to the Ca ion. These ultrafast charge transfer processes manifest as intense spectral features in the Auger spectrum, effectively masking the Ca$_{aq}^{3+}$(1s$^{-1}$) $\rightarrow$ Ca$_{aq}^{4+}$(2p$^{-2}$($^{1}$S)) Auger main line. These observations are similar to the Ca KLL Auger spectra of solid CaO, CaCO$_3$, and CaSO$_4$. In contrast, in K$^{+}$ solvated in water\citep{ceolin17:263003} charge transfer processes are much less intense so that the K$_{aq}^{2+}$(1s$^{-1}$) 
$\rightarrow$ K$_{aq}^{3+}$(2p$^{-2}$($^{1}$S)) Auger transition can be resolved.

A notable phenomenon observed in solvated Ca$^{2+}$ is a photon-energy dependent process manifested as an additional high-kinetic energy feature, which appears at photon energies of 4 eV above the ionization threshold. This feature is attributed to several Auger transitions emitting electrons of similar energies but originating from different processes. The first possibility is a charge transfer from a Cl$_{aq}^{-}$ ion in the second solvation shell to the Ca$^{3+}$ ion, which is suppressed as long as the photoelectron is still in the vicinity of the metal ion.  The second possibility is the Auger decay of Ca 1s$^{-1}$ photoelectron satellites, which are expected to begin at photon energies of 10 eV above threshold, and can contribute to the emission of Auger electrons in the energy region under discussion. A third  possible contribution comes from electron capture from the solution upon core ionization, which can occur when the photoelectron is sufficiently far from the ion, resulting in an Auger transition of the type Ca$_{aq}^{2+}$(1s$^{-1}$nl) $\rightarrow$ Ca$_{aq}^{3+}$(2p$^{-2}$($^1$D)nl), where nl is most likely 4s. The fact that different processes occurring at different photon energies lead to the emission of electrons of the same kinetic energy highlights the importance of recording Auger spectra at varying photon energies, especially those just above the ionization threshold. 

Our experiment also clearly demonstrates the presence of post-collision interaction (PCI) in the Auger spectra of aqueous Ca$^{2+}$, resulting in a PCI shift of about +1.2 eV at the threshold. Finally, the satellite structure of the Ca$^{3+}_{aq}$ (1s$^{-1}$) photoelectron spectrum is found to be dominated by structures, which are caused by electron energy loss processes of the photoelectron at the surrounding water molecules.  Since similar satellite structures are found for K$^{2+}_{aq}$ (1s$^{-1}$) and Cl$^{0}_{aq}$ (1s$^{-1}$) spectra, we expect that this finding is rather typical for core-hole photoelectron spectra of ions solvated in water. In general, such processes are also expected for Auger spectra, however, in the present case of Ca$^{2+}_{aq}$ such contributions are not observed due to much stronger and overlapping Auger transitions.

To gain deeper insight into these complex ultrafast charge-transfer processes in liquids, further measurements and calculations are needed. It would be particularly interesting to investigate which factors influence the intensity of the corresponding CT states — such as the metal ion charge, the species or concentration of the counterion, the solvent, or the solution’s pH.

These charge-transfer processes lead to the direct formation of highly reactive H$_2$O$^{+}$ radical cations, which play a crucial role in radiation damage\citep{ascenzi2023:24643}. Core ionization and core excitation in aqueous salt solutions initiate cascades of local and non-local electronic decay processes \citep{stumpf2016:237,gopakumar2023:1408}. The ultrafast formation of CT states will also impact the subsequent electronic decay cascade in aqueous solutions, affecting the energy distribution of slow electrons and the number of radical cations produced at the end of the cascade. 
\begin{acknowledgement}

Experiments were performed at the GALAXIES beamline, SOLEIL Synchrotron, France (Proposal No. 20160001). The authors are grateful to the SOLEIL staff for assistance during beamtime. TM thanks Dr.\ Iyas Ismail, Dr.\ Alexander Kuleff and Prof.\ Lorenz Cederbaum for valuable discussions.

\end{acknowledgement}

\begin{suppinfo}

This will usually read something like: ``Experimental procedures and
characterization data for all new compounds. The class will
automatically add a sentence pointing to the information on-line:

\end{suppinfo}

\bibliography{cacl2_bib}

\providecommand{\latin}[1]{#1}
\makeatletter
\providecommand{\doi}
  {\begingroup\let\do\@makeother\dospecials
  \catcode`\{=1 \catcode`\}=2 \doi@aux}
\providecommand{\doi@aux}[1]{\endgroup\texttt{#1}}
\makeatother
\providecommand*\mcitethebibliography{\thebibliography}
\csname @ifundefined\endcsname{endmcitethebibliography}
  {\let\endmcitethebibliography\endthebibliography}{}
\begin{mcitethebibliography}{65}
\providecommand*\natexlab[1]{#1}
\providecommand*\mciteSetBstSublistMode[1]{}
\providecommand*\mciteSetBstMaxWidthForm[2]{}
\providecommand*\mciteBstWouldAddEndPuncttrue
  {\def\EndOfBibitem{\unskip.}}
\providecommand*\mciteBstWouldAddEndPunctfalse
  {\let\EndOfBibitem\relax}
\providecommand*\mciteSetBstMidEndSepPunct[3]{}
\providecommand*\mciteSetBstSublistLabelBeginEnd[3]{}
\providecommand*\EndOfBibitem{}
\mciteSetBstSublistMode{f}
\mciteSetBstMaxWidthForm{subitem}{(\alph{mcitesubitemcount})}
\mciteSetBstSublistLabelBeginEnd
  {\mcitemaxwidthsubitemform\space}
  {\relax}
  {\relax}

\bibitem[Baird(2011)]{baird11:696}
Baird,~G.~S. Ionized calcium. \emph{Clinica Chimica Acta} \textbf{2011},
  \emph{412}, 696--701\relax
\mciteBstWouldAddEndPuncttrue
\mciteSetBstMidEndSepPunct{\mcitedefaultmidpunct}
{\mcitedefaultendpunct}{\mcitedefaultseppunct}\relax
\EndOfBibitem
\bibitem[C\'eolin \latin{et~al.}(2017)C\'eolin, Kryzhevoi, Nicolas, Pokapanich,
  Choksakulporn, Songsiriritthigul, Saisopa, Rattanachai, Utsumi, Palaudoux,
  \"Ohrwall, and Rueff]{ceolin17:263003}
C\'eolin,~D.; Kryzhevoi,~N.~V.; Nicolas,~C.; Pokapanich,~W.; Choksakulporn,~S.;
  Songsiriritthigul,~P.; Saisopa,~T.; Rattanachai,~Y.; Utsumi,~Y.;
  Palaudoux,~J.; \"Ohrwall,~G.; Rueff,~J.-P. Ultrafast Charge Transfer
  Processes Accompanying $KLL$ Auger Decay in Aqueous KCl Solution. \emph{Phys.
  Rev. Lett.} \textbf{2017}, \emph{119}, 263003\relax
\mciteBstWouldAddEndPuncttrue
\mciteSetBstMidEndSepPunct{\mcitedefaultmidpunct}
{\mcitedefaultendpunct}{\mcitedefaultseppunct}\relax
\EndOfBibitem
\bibitem[Dupuy \latin{et~al.}(2024)Dupuy, Buttersack, Trinter, Richter,
  Gholami, Bj{\"o}rneholm, Hergenhahn, Winter, and Bluhm]{dupuy24:6926}
Dupuy,~R.; Buttersack,~T.; Trinter,~F.; Richter,~C.; Gholami,~S.;
  Bj{\"o}rneholm,~O.; Hergenhahn,~U.; Winter,~B.; Bluhm,~H. The solvation shell
  probed by resonant intermolecular Coulombic decay. \emph{Nature
  Communications} \textbf{2024}, \emph{15}, 6926\relax
\mciteBstWouldAddEndPuncttrue
\mciteSetBstMidEndSepPunct{\mcitedefaultmidpunct}
{\mcitedefaultendpunct}{\mcitedefaultseppunct}\relax
\EndOfBibitem
\bibitem[Wörner \latin{et~al.}(2017)Wörner, Arrell, Banerji, Cannizzo,
  Chergui, Das, Hamm, Keller, Kraus, Liberatore, Lopez-Tarifa, Lucchini,
  Meuwly, Milne, Moser, Rothlisberger, Smolentsev, Teuscher, van Bokhoven, and
  Wenger]{woerner17:061508}
Wörner,~H.~J. \latin{et~al.}  Charge migration and charge transfer in
  molecular systems. \emph{Structural Dynamics} \textbf{2017}, \emph{4},
  061508\relax
\mciteBstWouldAddEndPuncttrue
\mciteSetBstMidEndSepPunct{\mcitedefaultmidpunct}
{\mcitedefaultendpunct}{\mcitedefaultseppunct}\relax
\EndOfBibitem
\bibitem[Bj\"orneholm \latin{et~al.}(1992)Bj\"orneholm, Nilsson, Sandell,
  Hernn\"as, and M{\aa}rtensson]{bjorneholm92:1892}
Bj\"orneholm,~O.; Nilsson,~A.; Sandell,~A.; Hernn\"as,~B.; M{\aa}rtensson,~N.
  Determination of time scales for charge-transfer screening in physisorbed
  molecules. \emph{Phys. Rev. Lett.} \textbf{1992}, \emph{68}, 1892--1895\relax
\mciteBstWouldAddEndPuncttrue
\mciteSetBstMidEndSepPunct{\mcitedefaultmidpunct}
{\mcitedefaultendpunct}{\mcitedefaultseppunct}\relax
\EndOfBibitem
\bibitem[F{\"o}hlisch \latin{et~al.}(2005)F{\"o}hlisch, Feulner, Hennies, Fink,
  Menzel, Sanchez-Portal, Echenique, and Wurth]{foehlisch05:373}
F{\"o}hlisch,~A.; Feulner,~P.; Hennies,~F.; Fink,~A.; Menzel,~D.;
  Sanchez-Portal,~D.; Echenique,~P.~M.; Wurth,~W. Direct observation of
  electron dynamics in the attosecond domain. \emph{Nature} \textbf{2005},
  \emph{436}, 373--376\relax
\mciteBstWouldAddEndPuncttrue
\mciteSetBstMidEndSepPunct{\mcitedefaultmidpunct}
{\mcitedefaultendpunct}{\mcitedefaultseppunct}\relax
\EndOfBibitem
\bibitem[Br\"uhwiler \latin{et~al.}(2002)Br\"uhwiler, Karis, and
  M\aa{}rtensson]{bruhwiler02:703}
Br\"uhwiler,~P.~A.; Karis,~O.; M\aa{}rtensson,~N. Charge-transfer dynamics
  studied using resonant core spectroscopies. \emph{Rev. Mod. Phys.}
  \textbf{2002}, \emph{74}, 703--740\relax
\mciteBstWouldAddEndPuncttrue
\mciteSetBstMidEndSepPunct{\mcitedefaultmidpunct}
{\mcitedefaultendpunct}{\mcitedefaultseppunct}\relax
\EndOfBibitem
\bibitem[Wurth and Menzel(2000)Wurth, and Menzel]{wurth00:141}
Wurth,~W.; Menzel,~D. Ultrafast electron dynamics at surfaces probed by
  resonant Auger spectroscopy. \emph{Chemical Physics} \textbf{2000},
  \emph{251}, 141--149\relax
\mciteBstWouldAddEndPuncttrue
\mciteSetBstMidEndSepPunct{\mcitedefaultmidpunct}
{\mcitedefaultendpunct}{\mcitedefaultseppunct}\relax
\EndOfBibitem
\bibitem[Miteva \latin{et~al.}(2018)Miteva, Kryzhevoi, Sisourat, Nicolas,
  Pokapanich, Saisopa, Songsiriritthigul, Rattanachai, Dreuw, Wenzel,
  Palaudoux, {\"O}hrwall, P{\"u}ttner, Cederbaum, Rueff, and
  C{\'e}olin]{miteva18:4457}
Miteva,~T. \latin{et~al.}  The All-Seeing Eye of Resonant Auger Electron
  Spectroscopy: A Study on Aqueous Solution Using Tender X-rays. \emph{The
  Journal of Physical Chemistry Letters} \textbf{2018}, \emph{9}, 4457--4462,
  PMID: 30020787\relax
\mciteBstWouldAddEndPuncttrue
\mciteSetBstMidEndSepPunct{\mcitedefaultmidpunct}
{\mcitedefaultendpunct}{\mcitedefaultseppunct}\relax
\EndOfBibitem
\bibitem[Nishikida and Ikeda(1978)Nishikida, and Ikeda]{nishikida78:49}
Nishikida,~S.; Ikeda,~S. Studies on the shake-up satellites of auger spectra
  for potassium and calcium compounds. \emph{Journal of Electron Spectroscopy
  and Related Phenomena} \textbf{1978}, \emph{13}, 49--58\relax
\mciteBstWouldAddEndPuncttrue
\mciteSetBstMidEndSepPunct{\mcitedefaultmidpunct}
{\mcitedefaultendpunct}{\mcitedefaultseppunct}\relax
\EndOfBibitem
\bibitem[Pokapanich \latin{et~al.}(2011)Pokapanich, Kryzhevoi, Ottosson,
  Svensson, Cederbaum, {\"O}hrwall, and Bj{\"o}rneholm]{pokapanich11:13430}
Pokapanich,~W.; Kryzhevoi,~N.~V.; Ottosson,~N.; Svensson,~S.; Cederbaum,~L.~S.;
  {\"O}hrwall,~G.; Bj{\"o}rneholm,~O. Ionic-charge dependence of the
  intermolecular coulombic decay time scale for aqueous ions probed by the
  core-hole clock. \emph{J Am Chem Soc} \textbf{2011}, \emph{133},
  13430--13436\relax
\mciteBstWouldAddEndPuncttrue
\mciteSetBstMidEndSepPunct{\mcitedefaultmidpunct}
{\mcitedefaultendpunct}{\mcitedefaultseppunct}\relax
\EndOfBibitem
\bibitem[Pokapanich \latin{et~al.}(2009)Pokapanich, Bergersen, Bradeanu,
  Marinho, Lindblad, Legendre, Rosso, Svensson, Bj{\"o}rneholm, Tchaplyguine,
  {\"O}hrwall, Kryzhevoi, and Cederbaum]{pokapanich09:7264}
Pokapanich,~W.; Bergersen,~H.; Bradeanu,~I.~L.; Marinho,~R. R.~T.;
  Lindblad,~A.; Legendre,~S.; Rosso,~A.; Svensson,~S.; Bj{\"o}rneholm,~O.;
  Tchaplyguine,~M.; {\"O}hrwall,~G.; Kryzhevoi,~N.~V.; Cederbaum,~L.~S. Auger
  Electron Spectroscopy as a Probe of the Solution of Aqueous Ions.
  \emph{Journal of the American Chemical Society} \textbf{2009}, \emph{131},
  7264--7271\relax
\mciteBstWouldAddEndPuncttrue
\mciteSetBstMidEndSepPunct{\mcitedefaultmidpunct}
{\mcitedefaultendpunct}{\mcitedefaultseppunct}\relax
\EndOfBibitem
\bibitem[Ottosson \latin{et~al.}(2012)Ottosson, Öhrwall, and
  Björneholm]{ottosson12:1}
Ottosson,~N.; Öhrwall,~G.; Björneholm,~O. Ultrafast charge delocalization
  dynamics in aqueous electrolytes: New insights from Auger electron
  spectroscopy. \emph{Chemical Physics Letters} \textbf{2012}, \emph{543},
  1--11\relax
\mciteBstWouldAddEndPuncttrue
\mciteSetBstMidEndSepPunct{\mcitedefaultmidpunct}
{\mcitedefaultendpunct}{\mcitedefaultseppunct}\relax
\EndOfBibitem
\bibitem[Simon \latin{et~al.}(2014)Simon, P{\"u}ttner, Marchenko, Guillemin,
  Kushawaha, Journel, Goldsztejn, Piancastelli, Ablett, Rueff, and
  C{\'e}olin]{simon14:4069}
Simon,~M.; P{\"u}ttner,~R.; Marchenko,~T.; Guillemin,~R.; Kushawaha,~R.~K.;
  Journel,~L.; Goldsztejn,~G.; Piancastelli,~M.~N.; Ablett,~J.~M.;
  Rueff,~J.-P.; C{\'e}olin,~D. Atomic Auger Doppler effects upon emission of
  fast photoelectrons. \emph{Nature Communications} \textbf{2014}, \emph{5},
  4069\relax
\mciteBstWouldAddEndPuncttrue
\mciteSetBstMidEndSepPunct{\mcitedefaultmidpunct}
{\mcitedefaultendpunct}{\mcitedefaultseppunct}\relax
\EndOfBibitem
\bibitem[Céolin \latin{et~al.}(2019)Céolin, Liu, da~Cruz, Ågren, Journel,
  Guillemin, Marchenko, Kushawaha, Piancastelli, Püttner, Simon, and
  Gel’mukhanov]{ceolin19:4877}
Céolin,~D.; Liu,~J.-C.; da~Cruz,~V.~V.; Ågren,~H.; Journel,~L.;
  Guillemin,~R.; Marchenko,~T.; Kushawaha,~R.~K.; Piancastelli,~M.~N.;
  Püttner,~R.; Simon,~M.; Gel’mukhanov,~F. Recoil-induced ultrafast
  molecular rotation probed by dynamical rotational Doppler effect.
  \emph{Proceedings of the National Academy of Sciences} \textbf{2019},
  \emph{116}, 4877--4882\relax
\mciteBstWouldAddEndPuncttrue
\mciteSetBstMidEndSepPunct{\mcitedefaultmidpunct}
{\mcitedefaultendpunct}{\mcitedefaultseppunct}\relax
\EndOfBibitem
\bibitem[Céolin \latin{et~al.}(2013)Céolin, Ablett, Prieur, Moreno, Rueff,
  Marchenko, Journel, Guillemin, Pilette, Marin, and Simon]{ceolin13:188}
Céolin,~D.; Ablett,~J.; Prieur,~D.; Moreno,~T.; Rueff,~J.-P.; Marchenko,~T.;
  Journel,~L.; Guillemin,~R.; Pilette,~B.; Marin,~T.; Simon,~M. Hard X-ray
  photoelectron spectroscopy on the GALAXIES beamline at the SOLEIL
  synchrotron. \emph{Journal of Electron Spectroscopy and Related Phenomena}
  \textbf{2013}, \emph{190}, 188--192, Recent advances in Hard X-ray
  Photoelectron Spectroscopy (HAXPES)\relax
\mciteBstWouldAddEndPuncttrue
\mciteSetBstMidEndSepPunct{\mcitedefaultmidpunct}
{\mcitedefaultendpunct}{\mcitedefaultseppunct}\relax
\EndOfBibitem
\bibitem[Rueff \latin{et~al.}(2015)Rueff, Ablett, C{\'{e}}olin, Prieur, Moreno,
  Bal{\'{e}}dent, Lassalle-Kaiser, Rault, Simon, and Shukla]{rueff2015:175}
Rueff,~J.-P.; Ablett,~J.~M.; C{\'{e}}olin,~D.; Prieur,~D.; Moreno,~T.;
  Bal{\'{e}}dent,~V.; Lassalle-Kaiser,~B.; Rault,~J.~E.; Simon,~M.; Shukla,~A.
  {The GALAXIES beamline at the SOLEIL synchrotron: inelastic X-ray scattering
  and photoelectron spectroscopy in the hard X-ray range}. \emph{Journal of
  Synchrotron Radiation} \textbf{2015}, \emph{22}, 175--179\relax
\mciteBstWouldAddEndPuncttrue
\mciteSetBstMidEndSepPunct{\mcitedefaultmidpunct}
{\mcitedefaultendpunct}{\mcitedefaultseppunct}\relax
\EndOfBibitem
\bibitem[Thürmer \latin{et~al.}(2021)Thürmer, Malerz, Trinter, Hergenhahn,
  Lee, Neumark, Meijer, Winter, and Wilkinson]{thurmer21:10558}
Thürmer,~S.; Malerz,~S.; Trinter,~F.; Hergenhahn,~U.; Lee,~C.; Neumark,~D.~M.;
  Meijer,~G.; Winter,~B.; Wilkinson,~I. Accurate vertical ionization energy and
  work function determinations of liquid water and aqueous solutions.
  \emph{Chem. Sci.} \textbf{2021}, \emph{12}, 10558--10582\relax
\mciteBstWouldAddEndPuncttrue
\mciteSetBstMidEndSepPunct{\mcitedefaultmidpunct}
{\mcitedefaultendpunct}{\mcitedefaultseppunct}\relax
\EndOfBibitem
\bibitem[Saisopa \latin{et~al.}(2020)Saisopa, Klaiphet, Songsiriritthigul,
  Pokapanich, Tangsukworakhun, Songsiriritthigul, Saiyasombat, Rattanachai,
  Yuzawa, Kosugi, and Céolin]{saisopa2020:146984}
Saisopa,~T.; Klaiphet,~K.; Songsiriritthigul,~P.; Pokapanich,~W.;
  Tangsukworakhun,~S.; Songsiriritthigul,~C.; Saiyasombat,~C.; Rattanachai,~Y.;
  Yuzawa,~H.; Kosugi,~N.; Céolin,~D. Investigation of solvated calcium
  dication structure in pure water, methanol, and ethanol solutions by means of
  K and L$_{2,3}$-edges X-ray absorption spectroscopy. \emph{Journal of
  Electron Spectroscopy and Related Phenomena} \textbf{2020}, \emph{244},
  146984\relax
\mciteBstWouldAddEndPuncttrue
\mciteSetBstMidEndSepPunct{\mcitedefaultmidpunct}
{\mcitedefaultendpunct}{\mcitedefaultseppunct}\relax
\EndOfBibitem
\bibitem[Megyes \latin{et~al.}(2004)Megyes, Grósz, Radnai, Bakó, and
  Pálinkás]{megyes04:7261}
Megyes,~T.; Grósz,~T.; Radnai,~T.; Bakó,~I.; Pálinkás,~G. Solvation of
  Calcium Ion in Polar Solvents: An X-ray Diffraction and ab Initio Study.
  \emph{The Journal of Physical Chemistry A} \textbf{2004}, \emph{108},
  7261--7271\relax
\mciteBstWouldAddEndPuncttrue
\mciteSetBstMidEndSepPunct{\mcitedefaultmidpunct}
{\mcitedefaultendpunct}{\mcitedefaultseppunct}\relax
\EndOfBibitem
\bibitem[Shao \latin{et~al.}(2015)Shao, Gan, Epifanovsky, Gilbert, Wormit,
  Kussmann, Lange, Behn, Deng, Feng, Ghosh, Goldey, Horn, Jacobson, Kaliman,
  Khaliullin, Kuś, Landau, Liu, Proynov, Rhee, Richard, Rohrdanz, Steele,
  Sundstrom, Woodcock~III, Zimmerman, Zuev, Albrecht, Alguire, Austin, Beran,
  Bernard, Berquist, Brandhorst, Bravaya, Brown, Casanova, Chang, Chen, Chien,
  Closser, Crittenden, Diedenhofen, DiStasio~Jr., Do, Dutoi, Edgar, Fatehi,
  Fusti-Molnar, Ghysels, Golubeva-Zadorozhnaya, Gomes, Hanson-Heine, Harbach,
  Hauser, Hohenstein, Holden, Jagau, Ji, Kaduk, Khistyaev, Kim, Jihan, King,
  Klunzinger, Kosenkov, Kowalczyk, Krauter, Lao, Laurent, Lawler, Levchenko,
  Lin, Liu, Livshits, Lochan, Luenser, Manohar, Manzer, Mao, Mardirossian,
  Marenich, Maurer, Mayhall, Neuscamman, Oana, Olivares-Amaya, O’Neill,
  Parkhill, Perrine, Peverati, Prociuk, Rehn, Rosta, Russ, Sharada, Sharma,
  Small, Sodt, Stein, Stück, Su, Thom, Tsuchimochi, Vanovschi, Vogt, Vydrov,
  Wang, Watson, Wenzel, White, Williams, Yang, Yeganeh, R., You, Zhang, Zhang,
  Zhao, Brooks, Chan, Chipman, Cramer, Goddard~III, Gordon, Hehre, Klamt,
  Schaefer~III, Schmidt, Sherrill, Truhlar, Warshel, Xu, Aspuru-Guzik, Baer,
  Bell, Besley, Chai, Dreuw, Dunietz, Furlani, Gwaltney, Hsu, Jung, Kong,
  Lambrecht, Liang, Ochsenfeld, Rassolov, Slipchenko, Subotnik, Voorhis,
  Herbert, Krylov, Gill, and Head-Gordon]{Shao17012015}
Shao,~Y. \latin{et~al.}  Advances in molecular quantum chemistry contained in
  the Q-Chem 4 program package. \emph{Molecular Physics} \textbf{2015},
  \emph{113}, 184--215\relax
\mciteBstWouldAddEndPuncttrue
\mciteSetBstMidEndSepPunct{\mcitedefaultmidpunct}
{\mcitedefaultendpunct}{\mcitedefaultseppunct}\relax
\EndOfBibitem
\bibitem[Jalilehvand \latin{et~al.}(2001)Jalilehvand, Sp{\aa}ngberg,
  Lindqvist-Reis, Hermansson, Persson, and Sandstr{\"o}m]{jalilehvand2001:431}
Jalilehvand,~F.; Sp{\aa}ngberg,~D.; Lindqvist-Reis,~P.; Hermansson,~K.;
  Persson,~I.; Sandstr{\"o}m,~M. Hydration of the Calcium Ion. An EXAFS,
  Large-Angle X-ray Scattering, and Molecular Dynamics Simulation Study.
  \emph{Journal of the American Chemical Society} \textbf{2001}, \emph{123},
  431--441, PMID: 11456545\relax
\mciteBstWouldAddEndPuncttrue
\mciteSetBstMidEndSepPunct{\mcitedefaultmidpunct}
{\mcitedefaultendpunct}{\mcitedefaultseppunct}\relax
\EndOfBibitem
\bibitem[D'Angelo \latin{et~al.}(2004)D'Angelo, Petit, and
  Pavel]{dangelo04:11857}
D'Angelo,~P.; Petit,~P.-E.; Pavel,~N.~V. Double-Electron Excitation Channels at
  the Ca$^{2+}$ K-Edge of Hydrated Calcium Ion. \emph{The Journal of Physical
  Chemistry B} \textbf{2004}, \emph{108}, 11857--11865\relax
\mciteBstWouldAddEndPuncttrue
\mciteSetBstMidEndSepPunct{\mcitedefaultmidpunct}
{\mcitedefaultendpunct}{\mcitedefaultseppunct}\relax
\EndOfBibitem
\bibitem[Tongraar \latin{et~al.}(2010)Tongraar, T-Thienprasert, Rujirawat, and
  Limpijumnong]{tongraar10:10876}
Tongraar,~A.; T-Thienprasert,~J.; Rujirawat,~S.; Limpijumnong,~S. Structure of
  the hydrated Ca$^{2+}$ and Cl$^-$: Combined X-ray absorption measurements and
  QM/MM MD simulations study. \emph{Phys. Chem. Chem. Phys.} \textbf{2010},
  \emph{12}, 10876--10887\relax
\mciteBstWouldAddEndPuncttrue
\mciteSetBstMidEndSepPunct{\mcitedefaultmidpunct}
{\mcitedefaultendpunct}{\mcitedefaultseppunct}\relax
\EndOfBibitem
\bibitem[Schirmer(1982)]{schirmer82:2395}
Schirmer,~J. Beyond the random-phase approximation: A new approximation scheme
  for the polarization propagator. \emph{Phys. Rev. A} \textbf{1982},
  \emph{26}, 2395--2416\relax
\mciteBstWouldAddEndPuncttrue
\mciteSetBstMidEndSepPunct{\mcitedefaultmidpunct}
{\mcitedefaultendpunct}{\mcitedefaultseppunct}\relax
\EndOfBibitem
\bibitem[Cederbaum \latin{et~al.}(1980)Cederbaum, Domcke, and
  Schirmer]{cederbaum80:206}
Cederbaum,~L.~S.; Domcke,~W.; Schirmer,~J. Many-body theory of core holes.
  \emph{Phys. Rev. A} \textbf{1980}, \emph{22}, 206--222\relax
\mciteBstWouldAddEndPuncttrue
\mciteSetBstMidEndSepPunct{\mcitedefaultmidpunct}
{\mcitedefaultendpunct}{\mcitedefaultseppunct}\relax
\EndOfBibitem
\bibitem[Barth and Cederbaum(1981)Barth, and Cederbaum]{cederbaum81:1038}
Barth,~A.; Cederbaum,~L.~S. Many-body theory of core-valence excitations.
  \emph{Phys. Rev. A} \textbf{1981}, \emph{23}, 1038--1061\relax
\mciteBstWouldAddEndPuncttrue
\mciteSetBstMidEndSepPunct{\mcitedefaultmidpunct}
{\mcitedefaultendpunct}{\mcitedefaultseppunct}\relax
\EndOfBibitem
\bibitem[Barth and Schirmer(1985)Barth, and Schirmer]{barth85:867}
Barth,~A.; Schirmer,~J. Theoretical core-level excitation spectra of N2 and CO
  by a new polarisation propagator method. \emph{Journal of Physics B: Atomic
  and Molecular Physics} \textbf{1985}, \emph{18}, 867\relax
\mciteBstWouldAddEndPuncttrue
\mciteSetBstMidEndSepPunct{\mcitedefaultmidpunct}
{\mcitedefaultendpunct}{\mcitedefaultseppunct}\relax
\EndOfBibitem
\bibitem[Wenzel \latin{et~al.}(2014)Wenzel, Wormit, and Dreuw]{wenzel2014:1900}
Wenzel,~J.; Wormit,~M.; Dreuw,~A. Calculating core-level excitations and x-ray
  absorption spectra of medium-sized closed-shell molecules with the
  algebraic-diagrammatic construction scheme for the polarization propagator.
  \emph{Journal of Computational Chemistry} \textbf{2014}, \emph{35},
  1900--1915\relax
\mciteBstWouldAddEndPuncttrue
\mciteSetBstMidEndSepPunct{\mcitedefaultmidpunct}
{\mcitedefaultendpunct}{\mcitedefaultseppunct}\relax
\EndOfBibitem
\bibitem[Wenzel \latin{et~al.}(2014)Wenzel, Wormit, and Dreuw]{wenzel2014:4583}
Wenzel,~J.; Wormit,~M.; Dreuw,~A. Calculating X-ray Absorption Spectra of
  Open-Shell Molecules with the Unrestricted Algebraic-Diagrammatic
  Construction Scheme for the Polarization Propagator. \emph{Journal of
  Chemical Theory and Computation} \textbf{2014}, \emph{10}, 4583--4598, PMID:
  26588152\relax
\mciteBstWouldAddEndPuncttrue
\mciteSetBstMidEndSepPunct{\mcitedefaultmidpunct}
{\mcitedefaultendpunct}{\mcitedefaultseppunct}\relax
\EndOfBibitem
\bibitem[Wormit \latin{et~al.}(2014)Wormit, Rehn, Harbach, Wenzel, Krauter,
  Epifanovsky, and Dreuw]{wormit14:774}
Wormit,~M.; Rehn,~D.~R.; Harbach,~P.~H.; Wenzel,~J.; Krauter,~C.~M.;
  Epifanovsky,~E.; Dreuw,~A. Investigating excited electronic states using the
  algebraic diagrammatic construction (ADC) approach of the polarisation
  propagator. \emph{Molecular Physics} \textbf{2014}, \emph{112},
  774--784\relax
\mciteBstWouldAddEndPuncttrue
\mciteSetBstMidEndSepPunct{\mcitedefaultmidpunct}
{\mcitedefaultendpunct}{\mcitedefaultseppunct}\relax
\EndOfBibitem
\bibitem[Krause and Oliver(1979)Krause, and Oliver]{krause79:329}
Krause,~M.~O.; Oliver,~J. Natural widths of atomic K and L levels, K$_{\alpha}$
  X‐ray lines and several KLL Auger lines. \emph{Journal of Physical and
  Chemical Reference Data} \textbf{1979}, \emph{8}, 329–338\relax
\mciteBstWouldAddEndPuncttrue
\mciteSetBstMidEndSepPunct{\mcitedefaultmidpunct}
{\mcitedefaultendpunct}{\mcitedefaultseppunct}\relax
\EndOfBibitem
\bibitem[Miteva \latin{et~al.}(2016)Miteva, Wenzel, Klaiman, Dreuw, and
  Gokhberg]{miteva2016:16671}
Miteva,~T.; Wenzel,~J.; Klaiman,~S.; Dreuw,~A.; Gokhberg,~K. X-Ray absorption
  spectra of microsolvated metal cations. \emph{Phys. Chem. Chem. Phys.}
  \textbf{2016}, \emph{18}, 16671--16681\relax
\mciteBstWouldAddEndPuncttrue
\mciteSetBstMidEndSepPunct{\mcitedefaultmidpunct}
{\mcitedefaultendpunct}{\mcitedefaultseppunct}\relax
\EndOfBibitem
\bibitem[Ivanic(2003)]{ivanic03:9364}
Ivanic,~J. Direct configuration interaction and multiconfigurational
  self-consistent-field method for multiple active spaces with variable
  occupations. I. Method. \emph{The Journal of Chemical Physics} \textbf{2003},
  \emph{119}, 9364--9376\relax
\mciteBstWouldAddEndPuncttrue
\mciteSetBstMidEndSepPunct{\mcitedefaultmidpunct}
{\mcitedefaultendpunct}{\mcitedefaultseppunct}\relax
\EndOfBibitem
\bibitem[Schmidt \latin{et~al.}(1993)Schmidt, Baldridge, Boatz, Elbert, Gordon,
  Jensen, Koseki, Matsunaga, Nguyen, Su, Windus, Dupuis, and
  Montgomery~Jr]{schmidt93:1347}
Schmidt,~M.~W.; Baldridge,~K.~K.; Boatz,~J.~A.; Elbert,~S.~T.; Gordon,~M.~S.;
  Jensen,~J.~H.; Koseki,~S.; Matsunaga,~N.; Nguyen,~K.~A.; Su,~S.;
  Windus,~T.~L.; Dupuis,~M.; Montgomery~Jr,~J.~A. General atomic and molecular
  electronic structure system. \emph{Journal of Computational Chemistry}
  \textbf{1993}, \emph{14}, 1347--1363\relax
\mciteBstWouldAddEndPuncttrue
\mciteSetBstMidEndSepPunct{\mcitedefaultmidpunct}
{\mcitedefaultendpunct}{\mcitedefaultseppunct}\relax
\EndOfBibitem
\bibitem[Pritchard \latin{et~al.}(2019)Pritchard, Altarawy, Didier, Gibson, and
  Windus]{pritchard2019:4814}
Pritchard,~B.~P.; Altarawy,~D.; Didier,~B.; Gibson,~T.~D.; Windus,~T.~L. New
  Basis Set Exchange: An Open, Up-to-Date Resource for the Molecular Sciences
  Community. \emph{Journal of Chemical Information and Modeling} \textbf{2019},
  \emph{59}, 4814--4820, PMID: 31600445\relax
\mciteBstWouldAddEndPuncttrue
\mciteSetBstMidEndSepPunct{\mcitedefaultmidpunct}
{\mcitedefaultendpunct}{\mcitedefaultseppunct}\relax
\EndOfBibitem
\bibitem[Blaudeau \latin{et~al.}(1997)Blaudeau, McGrath, Curtiss, and
  Radom]{blaudeau97:5016}
Blaudeau,~J.-P.; McGrath,~M.~P.; Curtiss,~L.~A.; Radom,~L. Extension of
  Gaussian-2 (G2) theory to molecules containing third-row atoms K and Ca.
  \emph{The Journal of Chemical Physics} \textbf{1997}, \emph{107},
  5016--5021\relax
\mciteBstWouldAddEndPuncttrue
\mciteSetBstMidEndSepPunct{\mcitedefaultmidpunct}
{\mcitedefaultendpunct}{\mcitedefaultseppunct}\relax
\EndOfBibitem
\bibitem[Winter \latin{et~al.}(2007)Winter, Hergenhahn, Faubel, Björneholm,
  and Hertel]{winter2007:094501}
Winter,~B.; Hergenhahn,~U.; Faubel,~M.; Björneholm,~O.; Hertel,~I.~V. Hydrogen
  bonding in liquid water probed by resonant Auger-electron spectroscopy.
  \emph{The Journal of Chemical Physics} \textbf{2007}, \emph{127},
  094501\relax
\mciteBstWouldAddEndPuncttrue
\mciteSetBstMidEndSepPunct{\mcitedefaultmidpunct}
{\mcitedefaultendpunct}{\mcitedefaultseppunct}\relax
\EndOfBibitem
\bibitem[Dingfelder \latin{et~al.}(1998)Dingfelder, Hantke, Inokuti, and
  Paretzke]{dingfelder98:1}
Dingfelder,~M.; Hantke,~D.; Inokuti,~M.; Paretzke,~H.~G. Electron
  inelastic-scattering cross sections in liquid water. \emph{Radiation Physics
  and Chemistry} \textbf{1998}, \emph{53}, 1--18\relax
\mciteBstWouldAddEndPuncttrue
\mciteSetBstMidEndSepPunct{\mcitedefaultmidpunct}
{\mcitedefaultendpunct}{\mcitedefaultseppunct}\relax
\EndOfBibitem
\bibitem[Emfietzoglou \latin{et~al.}(2003)Emfietzoglou, Karava, Papamichael,
  and Moscovitch]{emfiezoglou03:2355}
Emfietzoglou,~D.; Karava,~K.; Papamichael,~G.; Moscovitch,~M. Monte Carlo
  simulation of the energy loss of low-energy electrons in liquid water.
  \emph{Phys. Med. Biol.} \textbf{2003}, \emph{48}, 2355--2371\relax
\mciteBstWouldAddEndPuncttrue
\mciteSetBstMidEndSepPunct{\mcitedefaultmidpunct}
{\mcitedefaultendpunct}{\mcitedefaultseppunct}\relax
\EndOfBibitem
\bibitem[P\"uttner \latin{et~al.}(2020)P\"uttner, Holzhey, Hrast,
  \ifmmode~\check{Z}\else \v{Z}\fi{}itnik, Goldsztejn, Marchenko, Guillemin,
  Journel, Koulentianos, Travnikova, Zmerli, C\'eolin, Azuma, Kosugi, Lago,
  Piancastelli, and Simon]{puettner2020:052832}
P\"uttner,~R. \latin{et~al.}  Argon $KLL$ Auger spectrum: Initial states,
  core-hole lifetimes, shake, and knock-down processes. \emph{Phys. Rev. A}
  \textbf{2020}, \emph{102}, 052832\relax
\mciteBstWouldAddEndPuncttrue
\mciteSetBstMidEndSepPunct{\mcitedefaultmidpunct}
{\mcitedefaultendpunct}{\mcitedefaultseppunct}\relax
\EndOfBibitem
\bibitem[Li \latin{et~al.}(2025)Li, Jana, Garcia-Esparza, Li, Kaminsky, Hamlyn,
  Rajiv Ramanujam~Prabhakar, Harry A.~Atwater, Ager, Sokara, Yano, and
  Crumlin]{li2024:xxxx}
Li,~H.; Jana,~A.; Garcia-Esparza,~A.~T.; Li,~X.; Kaminsky,~C.~J.; Hamlyn,~R.;
  Rajiv Ramanujam~Prabhakar,~R.~R.; Harry A.~Atwater,~H.~A.; Ager,~J.~W.;
  Sokara,~D.; Yano,~J.; Crumlin,~E.~J. When Photoelectrons Meet Gas Molecules:
  Determining the Role of Inelastic Scattering in Ambient Pressure X-ray
  Photoelectron Spectroscopy. \emph{ACS Central Science} \textbf{2025},
  \emph{11}, 98--106\relax
\mciteBstWouldAddEndPuncttrue
\mciteSetBstMidEndSepPunct{\mcitedefaultmidpunct}
{\mcitedefaultendpunct}{\mcitedefaultseppunct}\relax
\EndOfBibitem
\bibitem[Muchov{\'a} \latin{et~al.}(2024)Muchov{\'a}, Gopakumar, Unger,
  {\"O}hrwall, C{\'e}olin, Trinter, Wilkinson, Chatzigeorgiou,
  Slav{\'i}{\v{c}}ek, Hergenhahn, Winter, Caleman, and
  Bj{\"o}rneholm]{muchova2024:8903}
Muchov{\'a},~E.; Gopakumar,~G.; Unger,~I.; {\"O}hrwall,~G.; C{\'e}olin,~D.;
  Trinter,~F.; Wilkinson,~I.; Chatzigeorgiou,~E.; Slav{\'i}{\v{c}}ek,~P.;
  Hergenhahn,~U.; Winter,~B.; Caleman,~C.; Bj{\"o}rneholm,~O. Attosecond
  formation of charge-transfer-to-solvent states of aqueous ions probed using
  the core-hole-clock technique. \emph{Nature Communications} \textbf{2024},
  \emph{15}, 8903\relax
\mciteBstWouldAddEndPuncttrue
\mciteSetBstMidEndSepPunct{\mcitedefaultmidpunct}
{\mcitedefaultendpunct}{\mcitedefaultseppunct}\relax
\EndOfBibitem
\bibitem[Arrio \latin{et~al.}(2000)Arrio, Rossano, Brouder, Galoisy, and
  Calas]{arrio2000:454}
Arrio,~M.-A.; Rossano,~S.; Brouder,~C.; Galoisy,~L.; Calas,~G. Calculation of
  multipole transitions at the Fe K pre-edge through p-d hybridization in the
  Ligand Field Multiplet model. \emph{Europhysics Letters} \textbf{2000},
  \emph{51}, 454\relax
\mciteBstWouldAddEndPuncttrue
\mciteSetBstMidEndSepPunct{\mcitedefaultmidpunct}
{\mcitedefaultendpunct}{\mcitedefaultseppunct}\relax
\EndOfBibitem
\bibitem[Martin-Diaconescu \latin{et~al.}(2015)Martin-Diaconescu, Gennari,
  Gerey, Tsui, Kanady, Tran, P{\'e}caut, Maganas, Krewald, Gour{\'e}, Duboc,
  Yano, Agapie, Collomb, and DeBeer]{martind2015:1283}
Martin-Diaconescu,~V.; Gennari,~M.; Gerey,~B.; Tsui,~E.; Kanady,~J.; Tran,~R.;
  P{\'e}caut,~J.; Maganas,~D.; Krewald,~V.; Gour{\'e},~E.; Duboc,~C.; Yano,~J.;
  Agapie,~T.; Collomb,~M.-N.; DeBeer,~S. Ca K-Edge XAS as a Probe of Calcium
  Centers in Complex Systems. \emph{Inorganic Chemistry} \textbf{2015},
  \emph{54}, 1283--1292, PMID: 25492398\relax
\mciteBstWouldAddEndPuncttrue
\mciteSetBstMidEndSepPunct{\mcitedefaultmidpunct}
{\mcitedefaultendpunct}{\mcitedefaultseppunct}\relax
\EndOfBibitem
\bibitem[Kramida \latin{et~al.}(2024)Kramida, {Yu.~Ralchenko}, Reader, and {and
  NIST ASD Team}]{NIST_ASD}
Kramida,~A.; {Yu.~Ralchenko}; Reader,~J.; {and NIST ASD Team} {NIST Atomic
  Spectra Database (ver. 5.12), [Online]. Available:
  {\tt{https://physics.nist.gov/asd}} [2016, January 31]. National Institute of
  Standards and Technology, Gaithersburg, MD.}, 2024\relax
\mciteBstWouldAddEndPuncttrue
\mciteSetBstMidEndSepPunct{\mcitedefaultmidpunct}
{\mcitedefaultendpunct}{\mcitedefaultseppunct}\relax
\EndOfBibitem
\bibitem[Fulton \latin{et~al.}(2003)Fulton, Heald, Badyal, and
  Simonson]{fulton2003:4688}
Fulton,~J.~L.; Heald,~S.~M.; Badyal,~Y.~S.; Simonson,~J.~M. Understanding the
  Effects of Concentration on the Solvation Structure of Ca$^{2+}$ in Aqueous
  Solution. I: The Perspective on Local Structure from EXAFS and XANES.
  \emph{The Journal of Physical Chemistry A} \textbf{2003}, \emph{107},
  4688--4696\relax
\mciteBstWouldAddEndPuncttrue
\mciteSetBstMidEndSepPunct{\mcitedefaultmidpunct}
{\mcitedefaultendpunct}{\mcitedefaultseppunct}\relax
\EndOfBibitem
\bibitem[Fulton \latin{et~al.}(2006)Fulton, Chen, Heald, and
  Balasubramanian]{fulton2006:094507}
Fulton,~J.~L.; Chen,~Y.; Heald,~S.~M.; Balasubramanian,~M. Hydration and
  contact ion pairing of Ca$^{2+}$ with Cl$^-$ in supercritical aqueous
  solution. \emph{The Journal of Chemical Physics} \textbf{2006}, \emph{125},
  094507\relax
\mciteBstWouldAddEndPuncttrue
\mciteSetBstMidEndSepPunct{\mcitedefaultmidpunct}
{\mcitedefaultendpunct}{\mcitedefaultseppunct}\relax
\EndOfBibitem
\bibitem[Kav\ifmmode \check{c}\else \v{c}\fi{}i\ifmmode~\check{c}\else
  \v{c}\fi{} \latin{et~al.}(2009)Kav\ifmmode \check{c}\else
  \v{c}\fi{}i\ifmmode~\check{c}\else \v{c}\fi{}, \ifmmode~\check{Z}\else
  \v{Z}\fi{}itnik, Bu\ifmmode~\check{c}\else \v{c}\fi{}ar,
  Miheli\ifmmode~\check{c}\else \v{c}\fi{}, \ifmmode~\check{S}\else
  \v{S}\fi{}tuhec, Szlachetko, Cao, Alonso~Mori, and
  Glatzel]{kavcic2009:143001}
Kav\ifmmode \check{c}\else \v{c}\fi{}i\ifmmode~\check{c}\else \v{c}\fi{},~M.;
  \ifmmode~\check{Z}\else \v{Z}\fi{}itnik,~M.; Bu\ifmmode~\check{c}\else
  \v{c}\fi{}ar,~K.; Miheli\ifmmode~\check{c}\else \v{c}\fi{},~A.;
  \ifmmode~\check{S}\else \v{S}\fi{}tuhec,~M.; Szlachetko,~J.; Cao,~W.;
  Alonso~Mori,~R.; Glatzel,~P. Separation of Two-Electron Photoexcited Atomic
  Processes near the Inner-Shell Threshold. \emph{Phys. Rev. Lett.}
  \textbf{2009}, \emph{102}, 143001\relax
\mciteBstWouldAddEndPuncttrue
\mciteSetBstMidEndSepPunct{\mcitedefaultmidpunct}
{\mcitedefaultendpunct}{\mcitedefaultseppunct}\relax
\EndOfBibitem
\bibitem[{Sukhorukov, V.L.} \latin{et~al.}(1987){Sukhorukov, V.L.}, {Hopersky,
  A.N.}, {Petrov, I.D.}, {Yavna, V.A.}, and {Demekhin,
  V.F.}]{sukhorukov87:1677}
{Sukhorukov, V.L.}; {Hopersky, A.N.}; {Petrov, I.D.}; {Yavna, V.A.}; {Demekhin,
  V.F.} Double photoexcitation processes at the near K-edge region of Ne, Na
  and Ar. \emph{J. Phys. France} \textbf{1987}, \emph{48}, 1677--1683\relax
\mciteBstWouldAddEndPuncttrue
\mciteSetBstMidEndSepPunct{\mcitedefaultmidpunct}
{\mcitedefaultendpunct}{\mcitedefaultseppunct}\relax
\EndOfBibitem
\bibitem[\ifmmode~\check{Z}\else \v{Z}\fi{}itnik
  \latin{et~al.}(2023)\ifmmode~\check{Z}\else \v{Z}\fi{}itnik, Hrast,
  Miheli\ifmmode~\check{c}\else \v{c}\fi{}, Bu\ifmmode~\check{c}\else
  \v{c}\fi{}ar, Turn\ifmmode~\check{s}\else \v{s}\fi{}ek, P\"uttner,
  Goldsztejn, Marchenko, Guillemin, Journel, Travnikova, Ismail, Piancastelli,
  Simon, Ceolin, and Kav\ifmmode \check{c}\else
  \v{c}\fi{}i\ifmmode~\check{c}\else \v{c}\fi{}]{zitnik2023:053113}
\ifmmode~\check{Z}\else \v{Z}\fi{}itnik,~M. \latin{et~al.}  Auger decay of
  $1{s}^{\ensuremath{-}1}3{p}^{\ensuremath{-}1}nl{n}^{\ensuremath{'}}{l}^{\ensuremath{'}}$
  doubly excited states in Ar. \emph{Phys. Rev. A} \textbf{2023}, \emph{108},
  053113\relax
\mciteBstWouldAddEndPuncttrue
\mciteSetBstMidEndSepPunct{\mcitedefaultmidpunct}
{\mcitedefaultendpunct}{\mcitedefaultseppunct}\relax
\EndOfBibitem
\bibitem[Gel'mukhanov and \AA{}gren(1996)Gel'mukhanov, and
  \AA{}gren]{gelmukhanov96:3960}
Gel'mukhanov,~F.; \AA{}gren,~H. Raman, non-Raman, and anti-Raman dispersion in
  resonant x-ray scattering spectra of molecules. \emph{Phys. Rev. A}
  \textbf{1996}, \emph{54}, 3960--3970\relax
\mciteBstWouldAddEndPuncttrue
\mciteSetBstMidEndSepPunct{\mcitedefaultmidpunct}
{\mcitedefaultendpunct}{\mcitedefaultseppunct}\relax
\EndOfBibitem
\bibitem[Kukk \latin{et~al.}(1997)Kukk, Aksela, Kivim\"aki, Jauhiainen,
  N\~ommiste, and Aksela]{kukk97:1481}
Kukk,~E.; Aksela,~H.; Kivim\"aki,~A.; Jauhiainen,~J.; N\~ommiste,~E.;
  Aksela,~S. Electronic-state lifetime interference in the resonant Auger decay
  of krypton. \emph{Phys. Rev. A} \textbf{1997}, \emph{56}, 1481--1485\relax
\mciteBstWouldAddEndPuncttrue
\mciteSetBstMidEndSepPunct{\mcitedefaultmidpunct}
{\mcitedefaultendpunct}{\mcitedefaultseppunct}\relax
\EndOfBibitem
\bibitem[Asplund \latin{et~al.}(1977)Asplund, Kelfve, Blomster, Siegbahn, and
  Siegbahn]{asplund77:268}
Asplund,~L.; Kelfve,~P.; Blomster,~B.; Siegbahn,~H.; Siegbahn,~K. Argon KLL and
  KLM Auger Electron Spectra. \emph{Physica Scripta} \textbf{1977}, \emph{16},
  268\relax
\mciteBstWouldAddEndPuncttrue
\mciteSetBstMidEndSepPunct{\mcitedefaultmidpunct}
{\mcitedefaultendpunct}{\mcitedefaultseppunct}\relax
\EndOfBibitem
\bibitem[Velasquez \latin{et~al.}(2023)Velasquez, Travnikova, Guillemin,
  Ismail, Journel, Martins, Koulentianos, C\'eolin, Fillaud, Rocco, P\"uttner,
  Piancastelli, Simon, Sheinerman, Gerchikov, and
  Marchenko]{velsquez2023:013048}
Velasquez,~N. \latin{et~al.}  Generalization of the post-collision interaction
  effect from gas-phase to solid-state systems demonstrated in thiophene and
  its polymers. \emph{Phys. Rev. Res.} \textbf{2023}, \emph{5}, 013048\relax
\mciteBstWouldAddEndPuncttrue
\mciteSetBstMidEndSepPunct{\mcitedefaultmidpunct}
{\mcitedefaultendpunct}{\mcitedefaultseppunct}\relax
\EndOfBibitem
\bibitem[Todorova \latin{et~al.}(2008)Todorova, H{\"u}nenberger, and
  Hutter]{todorova2008:779}
Todorova,~T.; H{\"u}nenberger,~P.~H.; Hutter,~J. Car–Parrinello Molecular
  Dynamics Simulations of CaCl$_2$ Aqueous Solutions. \emph{Journal of Chemical
  Theory and Computation} \textbf{2008}, \emph{4}, 779--789, PMID:
  26621092\relax
\mciteBstWouldAddEndPuncttrue
\mciteSetBstMidEndSepPunct{\mcitedefaultmidpunct}
{\mcitedefaultendpunct}{\mcitedefaultseppunct}\relax
\EndOfBibitem
\bibitem[Friesen \latin{et~al.}(2019)Friesen, Hefter, and
  Buchner]{friesen2019:891}
Friesen,~S.; Hefter,~G.; Buchner,~R. Cation Hydration and Ion Pairing in
  Aqueous Solutions of MgCl$_2$ and CaCl$_2$. \emph{The Journal of Physical
  Chemistry B} \textbf{2019}, \emph{123}, 891--900\relax
\mciteBstWouldAddEndPuncttrue
\mciteSetBstMidEndSepPunct{\mcitedefaultmidpunct}
{\mcitedefaultendpunct}{\mcitedefaultseppunct}\relax
\EndOfBibitem
\bibitem[Marcus(1988)]{markus88:1475}
Marcus,~Y. Ionic radii in aqueous solutions. \emph{Chem. Rev.} \textbf{1988},
  \emph{88}, 1475–1498\relax
\mciteBstWouldAddEndPuncttrue
\mciteSetBstMidEndSepPunct{\mcitedefaultmidpunct}
{\mcitedefaultendpunct}{\mcitedefaultseppunct}\relax
\EndOfBibitem
\bibitem[Sinha and Antony(2021)Sinha, and Antony]{sinha2021:5479}
Sinha,~N.; Antony,~B. Mean Free Paths and Cross Sections for Electron
  Scattering from Liquid Water. \emph{The Journal of Physical Chemistry B}
  \textbf{2021}, \emph{125}, 5479--5488, PMID: 34014676\relax
\mciteBstWouldAddEndPuncttrue
\mciteSetBstMidEndSepPunct{\mcitedefaultmidpunct}
{\mcitedefaultendpunct}{\mcitedefaultseppunct}\relax
\EndOfBibitem
\bibitem[Schild \latin{et~al.}(2020)Schild, Peper, Perry, Rattenbacher, and
  W{\"o}rner]{schild2020:1128}
Schild,~A.; Peper,~M.; Perry,~C.; Rattenbacher,~D.; W{\"o}rner,~H.~J.
  Alternative Approach for the Determination of Mean Free Paths of Electron
  Scattering in Liquid Water Based on Experimental Data. \emph{The Journal of
  Physical Chemistry Letters} \textbf{2020}, \emph{11}, 1128--1134, PMID:
  31928019\relax
\mciteBstWouldAddEndPuncttrue
\mciteSetBstMidEndSepPunct{\mcitedefaultmidpunct}
{\mcitedefaultendpunct}{\mcitedefaultseppunct}\relax
\EndOfBibitem
\bibitem[Jahnke \latin{et~al.}(2020)Jahnke, Hergenhahn, Winter, D{\"o}rner,
  Fr{\"u}hling, Demekhin, Gokhberg, Cederbaum, Ehresmann, Knie, and
  Dreuw]{jahnke2020:11295}
Jahnke,~T.; Hergenhahn,~U.; Winter,~B.; D{\"o}rner,~R.; Fr{\"u}hling,~U.;
  Demekhin,~P.~V.; Gokhberg,~K.; Cederbaum,~L.~S.; Ehresmann,~A.; Knie,~A.;
  Dreuw,~A. Interatomic and Intermolecular Coulombic Decay. \emph{Chemical
  Reviews} \textbf{2020}, \emph{120}, 11295--11369, PMID: 33035051\relax
\mciteBstWouldAddEndPuncttrue
\mciteSetBstMidEndSepPunct{\mcitedefaultmidpunct}
{\mcitedefaultendpunct}{\mcitedefaultseppunct}\relax
\EndOfBibitem
\bibitem[Ascenzi \latin{et~al.}(2023)Ascenzi, Erdmann, Bolognesi, Avaldi,
  Castrovilli, Thissen, Romanzin, Alcaraz, Rabadan, Mendez, Díaz-Tendero, and
  Cartoni]{ascenzi2023:24643}
Ascenzi,~D.; Erdmann,~E.; Bolognesi,~P.; Avaldi,~L.; Castrovilli,~M.~C.;
  Thissen,~R.; Romanzin,~C.; Alcaraz,~C.; Rabadan,~I.; Mendez,~L.;
  Díaz-Tendero,~S.; Cartoni,~A. H$_2$O$^+$ and OH$^+$ reactivity versus furan:
  experimental low energy absolute cross sections for modeling radiation
  damage. \emph{Phys. Chem. Chem. Phys.} \textbf{2023}, \emph{25},
  24643--24656\relax
\mciteBstWouldAddEndPuncttrue
\mciteSetBstMidEndSepPunct{\mcitedefaultmidpunct}
{\mcitedefaultendpunct}{\mcitedefaultseppunct}\relax
\EndOfBibitem
\bibitem[Stumpf \latin{et~al.}(2016)Stumpf, Gokhberg, and
  Cederbaum]{stumpf2016:237}
Stumpf,~V.; Gokhberg,~K.; Cederbaum,~L.~S. The role of metal ions in
  X-ray-induced photochemistry. \emph{Nature Chemistry} \textbf{2016},
  \emph{8}, 237--241\relax
\mciteBstWouldAddEndPuncttrue
\mciteSetBstMidEndSepPunct{\mcitedefaultmidpunct}
{\mcitedefaultendpunct}{\mcitedefaultseppunct}\relax
\EndOfBibitem
\bibitem[Gopakumar \latin{et~al.}(2023)Gopakumar, Unger, Slav{\'i}{\v{c}}ek,
  Hergenhahn, {\"O}hrwall, Malerz, C{\'e}olin, Trinter, Winter, Wilkinson,
  Caleman, Muchov{\'a}, and Bj{\"o}rneholm]{gopakumar2023:1408}
Gopakumar,~G.; Unger,~I.; Slav{\'i}{\v{c}}ek,~P.; Hergenhahn,~U.;
  {\"O}hrwall,~G.; Malerz,~S.; C{\'e}olin,~D.; Trinter,~F.; Winter,~B.;
  Wilkinson,~I.; Caleman,~C.; Muchov{\'a},~E.; Bj{\"o}rneholm,~O. Radiation
  damage by extensive local water ionization from two-step
  electron-transfer-mediated decay of solvated ions. \emph{Nature Chemistry}
  \textbf{2023}, \emph{15}, 1408--1414\relax
\mciteBstWouldAddEndPuncttrue
\mciteSetBstMidEndSepPunct{\mcitedefaultmidpunct}
{\mcitedefaultendpunct}{\mcitedefaultseppunct}\relax
\EndOfBibitem
\end{mcitethebibliography}


\providecommand{\latin}[1]{#1}
\makeatletter
\providecommand{\doi}
  {\begingroup\let\do\@makeother\dospecials
  \catcode`\{=1 \catcode`\}=2 \doi@aux}
\providecommand{\doi@aux}[1]{\endgroup\texttt{#1}}
\makeatother
\providecommand*\mcitethebibliography{\thebibliography}
\csname @ifundefined\endcsname{endmcitethebibliography}
  {\let\endmcitethebibliography\endthebibliography}{}
\begin{mcitethebibliography}{15}
\providecommand*\natexlab[1]{#1}
\providecommand*\mciteSetBstSublistMode[1]{}
\providecommand*\mciteSetBstMaxWidthForm[2]{}
\providecommand*\mciteBstWouldAddEndPuncttrue
  {\def\EndOfBibitem{\unskip.}}
\providecommand*\mciteBstWouldAddEndPunctfalse
  {\let\EndOfBibitem\relax}
\providecommand*\mciteSetBstMidEndSepPunct[3]{}
\providecommand*\mciteSetBstSublistLabelBeginEnd[3]{}
\providecommand*\EndOfBibitem{}
\mciteSetBstSublistMode{f}
\mciteSetBstMaxWidthForm{subitem}{(\alph{mcitesubitemcount})}
\mciteSetBstSublistLabelBeginEnd
  {\mcitemaxwidthsubitemform\space}
  {\relax}
  {\relax}

\bibitem[Krause and Oliver(1979)Krause, and Oliver]{krause79:329}
Krause,~M.~O.; Oliver,~J. Natural widths of atomic K and L levels, K$_{\alpha}$
  X‐ray lines and several KLL Auger lines. \emph{Journal of Physical and
  Chemical Reference Data} \textbf{1979}, \emph{8}, 329–338\relax
\mciteBstWouldAddEndPuncttrue
\mciteSetBstMidEndSepPunct{\mcitedefaultmidpunct}
{\mcitedefaultendpunct}{\mcitedefaultseppunct}\relax
\EndOfBibitem
\bibitem[Jalilehvand \latin{et~al.}(2001)Jalilehvand, Sp{\aa}ngberg,
  Lindqvist-Reis, Hermansson, Persson, and Sandstr{\"o}m]{jalilehvand2001:431}
Jalilehvand,~F.; Sp{\aa}ngberg,~D.; Lindqvist-Reis,~P.; Hermansson,~K.;
  Persson,~I.; Sandstr{\"o}m,~M. Hydration of the Calcium Ion. An EXAFS,
  Large-Angle X-ray Scattering, and Molecular Dynamics Simulation Study.
  \emph{Journal of the American Chemical Society} \textbf{2001}, \emph{123},
  431--441, PMID: 11456545\relax
\mciteBstWouldAddEndPuncttrue
\mciteSetBstMidEndSepPunct{\mcitedefaultmidpunct}
{\mcitedefaultendpunct}{\mcitedefaultseppunct}\relax
\EndOfBibitem
\bibitem[D'Angelo \latin{et~al.}(2004)D'Angelo, Petit, and
  Pavel]{dangelo04:11857}
D'Angelo,~P.; Petit,~P.-E.; Pavel,~N.~V. Double-Electron Excitation Channels at
  the Ca$^{2+}$ K-Edge of Hydrated Calcium Ion. \emph{The Journal of Physical
  Chemistry B} \textbf{2004}, \emph{108}, 11857--11865\relax
\mciteBstWouldAddEndPuncttrue
\mciteSetBstMidEndSepPunct{\mcitedefaultmidpunct}
{\mcitedefaultendpunct}{\mcitedefaultseppunct}\relax
\EndOfBibitem
\bibitem[Tongraar \latin{et~al.}(2010)Tongraar, T-Thienprasert, Rujirawat, and
  Limpijumnong]{tongraar10:10876}
Tongraar,~A.; T-Thienprasert,~J.; Rujirawat,~S.; Limpijumnong,~S. Structure of
  the hydrated Ca$^{2+}$ and Cl$^-$: Combined X-ray absorption measurements and
  QM/MM MD simulations study. \emph{Phys. Chem. Chem. Phys.} \textbf{2010},
  \emph{12}, 10876--10887\relax
\mciteBstWouldAddEndPuncttrue
\mciteSetBstMidEndSepPunct{\mcitedefaultmidpunct}
{\mcitedefaultendpunct}{\mcitedefaultseppunct}\relax
\EndOfBibitem
\bibitem[Megyes \latin{et~al.}(2004)Megyes, Grósz, Radnai, Bakó, and
  Pálinkás]{megyes04:7261}
Megyes,~T.; Grósz,~T.; Radnai,~T.; Bakó,~I.; Pálinkás,~G. Solvation of
  Calcium Ion in Polar Solvents: An X-ray Diffraction and ab Initio Study.
  \emph{The Journal of Physical Chemistry A} \textbf{2004}, \emph{108},
  7261--7271\relax
\mciteBstWouldAddEndPuncttrue
\mciteSetBstMidEndSepPunct{\mcitedefaultmidpunct}
{\mcitedefaultendpunct}{\mcitedefaultseppunct}\relax
\EndOfBibitem
\bibitem[C\'eolin \latin{et~al.}(2017)C\'eolin, Kryzhevoi, Nicolas, Pokapanich,
  Choksakulporn, Songsiriritthigul, Saisopa, Rattanachai, Utsumi, Palaudoux,
  \"Ohrwall, and Rueff]{ceolin17:263003}
C\'eolin,~D.; Kryzhevoi,~N.~V.; Nicolas,~C.; Pokapanich,~W.; Choksakulporn,~S.;
  Songsiriritthigul,~P.; Saisopa,~T.; Rattanachai,~Y.; Utsumi,~Y.;
  Palaudoux,~J.; \"Ohrwall,~G.; Rueff,~J.-P. Ultrafast Charge Transfer
  Processes Accompanying $KLL$ Auger Decay in Aqueous KCl Solution. \emph{Phys.
  Rev. Lett.} \textbf{2017}, \emph{119}, 263003\relax
\mciteBstWouldAddEndPuncttrue
\mciteSetBstMidEndSepPunct{\mcitedefaultmidpunct}
{\mcitedefaultendpunct}{\mcitedefaultseppunct}\relax
\EndOfBibitem
\bibitem[P\"uttner \latin{et~al.}(2020)P\"uttner, Holzhey, Hrast,
  \ifmmode~\check{Z}\else \v{Z}\fi{}itnik, Goldsztejn, Marchenko, Guillemin,
  Journel, Koulentianos, Travnikova, Zmerli, C\'eolin, Azuma, Kosugi, Lago,
  Piancastelli, and Simon]{puettner2020:052832}
P\"uttner,~R. \latin{et~al.}  Argon $KLL$ Auger spectrum: Initial states,
  core-hole lifetimes, shake, and knock-down processes. \emph{Phys. Rev. A}
  \textbf{2020}, \emph{102}, 052832\relax
\mciteBstWouldAddEndPuncttrue
\mciteSetBstMidEndSepPunct{\mcitedefaultmidpunct}
{\mcitedefaultendpunct}{\mcitedefaultseppunct}\relax
\EndOfBibitem
\bibitem[Nishikida and Ikeda(1978)Nishikida, and Ikeda]{nishikida78:49}
Nishikida,~S.; Ikeda,~S. Studies on the shake-up satellites of auger spectra
  for potassium and calcium compounds. \emph{Journal of Electron Spectroscopy
  and Related Phenomena} \textbf{1978}, \emph{13}, 49--58\relax
\mciteBstWouldAddEndPuncttrue
\mciteSetBstMidEndSepPunct{\mcitedefaultmidpunct}
{\mcitedefaultendpunct}{\mcitedefaultseppunct}\relax
\EndOfBibitem
\bibitem[Winter \latin{et~al.}(2005)Winter, Weber, Hertel, Faubel, Jungwirth,
  Brown, and Bradforth]{winter2005:7203}
Winter,~B.; Weber,~R.; Hertel,~I.~V.; Faubel,~M.; Jungwirth,~P.; Brown,~E.~C.;
  Bradforth,~S.~E. Electron Binding Energies of Aqueous Alkali and Halide Ions:
  EUV Photoelectron Spectroscopy of Liquid Solutions and Combined Ab Initio and
  Molecular Dynamics Calculations. \emph{J. Am. Chem. Soc.} \textbf{2005},
  \emph{127}, 7203--7214\relax
\mciteBstWouldAddEndPuncttrue
\mciteSetBstMidEndSepPunct{\mcitedefaultmidpunct}
{\mcitedefaultendpunct}{\mcitedefaultseppunct}\relax
\EndOfBibitem
\bibitem[Kurahashi \latin{et~al.}(2014)Kurahashi, Karashima, Tang, Horio,
  Abulimiti, Suzuki, Ogi, Oura, and Suzuki]{kurahashi2014:174506}
Kurahashi,~N.; Karashima,~S.; Tang,~Y.; Horio,~T.; Abulimiti,~B.;
  Suzuki,~Y.-I.; Ogi,~Y.~O.; Oura,~M.~O.; Suzuki,~T. Photoelectron spectroscopy
  of aqueous solutions: Streaming potentials of NaX (X = Cl, Br, and I)
  solutions and electron binding energies of liquid water and X$^-$. \emph{J.
  Chem. Phys.} \textbf{2014}, \emph{140}, 174506\relax
\mciteBstWouldAddEndPuncttrue
\mciteSetBstMidEndSepPunct{\mcitedefaultmidpunct}
{\mcitedefaultendpunct}{\mcitedefaultseppunct}\relax
\EndOfBibitem
\bibitem[Yarzhemsky and Amusia(2016)Yarzhemsky, and
  Amusia]{yarzhemsky2016:063406}
Yarzhemsky,~V.~G.; Amusia,~M.~Y. Calculation of Ar photoelectron satellites in
  the hard-x-ray region. \emph{Phys. Rev. A} \textbf{2016}, \emph{93},
  063406\relax
\mciteBstWouldAddEndPuncttrue
\mciteSetBstMidEndSepPunct{\mcitedefaultmidpunct}
{\mcitedefaultendpunct}{\mcitedefaultseppunct}\relax
\EndOfBibitem
\bibitem[Dyall(1983)]{dyall1983:3137}
Dyall,~K. Shake theory predictions of excited-state populations following 1s
  ionisation in argon. \emph{Journal of Physics B: Atomic and Molecular
  Physics} \textbf{1983}, \emph{16}, 3137–3147\relax
\mciteBstWouldAddEndPuncttrue
\mciteSetBstMidEndSepPunct{\mcitedefaultmidpunct}
{\mcitedefaultendpunct}{\mcitedefaultseppunct}\relax
\EndOfBibitem
\bibitem[De~Fanis \latin{et~al.}(2004)De~Fanis, Saito, Okada, Machida, Koyano,
  Cassimi, D\"{o}rner, Pavlychev, and Ueda]{defanis2004:265}
De~Fanis,~A.; Saito,~N.; Okada,~K.; Machida,~M.; Koyano,~I.; Cassimi,~A.;
  D\"{o}rner,~R.; Pavlychev,~A.; Ueda,~K. Satellite excitations due to internal
  inelastic scattering in the K-shell photoemission from CO$_2$. \emph{Journal
  of Electron Spectroscopy and Related Phenomena} \textbf{2004},
  \emph{137–140}, 265--269\relax
\mciteBstWouldAddEndPuncttrue
\mciteSetBstMidEndSepPunct{\mcitedefaultmidpunct}
{\mcitedefaultendpunct}{\mcitedefaultseppunct}\relax
\EndOfBibitem
\bibitem[H\"{u}fner(2003)]{huefner2003}
H\"{u}fner,~S. \emph{Photoelectron Spectroscopy, Principles and Applications,
  Thrid Edition}; Springer: Berlin, 2003\relax
\mciteBstWouldAddEndPuncttrue
\mciteSetBstMidEndSepPunct{\mcitedefaultmidpunct}
{\mcitedefaultendpunct}{\mcitedefaultseppunct}\relax
\EndOfBibitem
\end{mcitethebibliography}

\end{document}



\section{Theoretical X-ray absorption spectra and radial density distributions}

\begin{figure}[H]
\renewcommand{\thefigure}{SI\arabic{figure}}
\centering
    \includegraphics[width=0.62\linewidth]{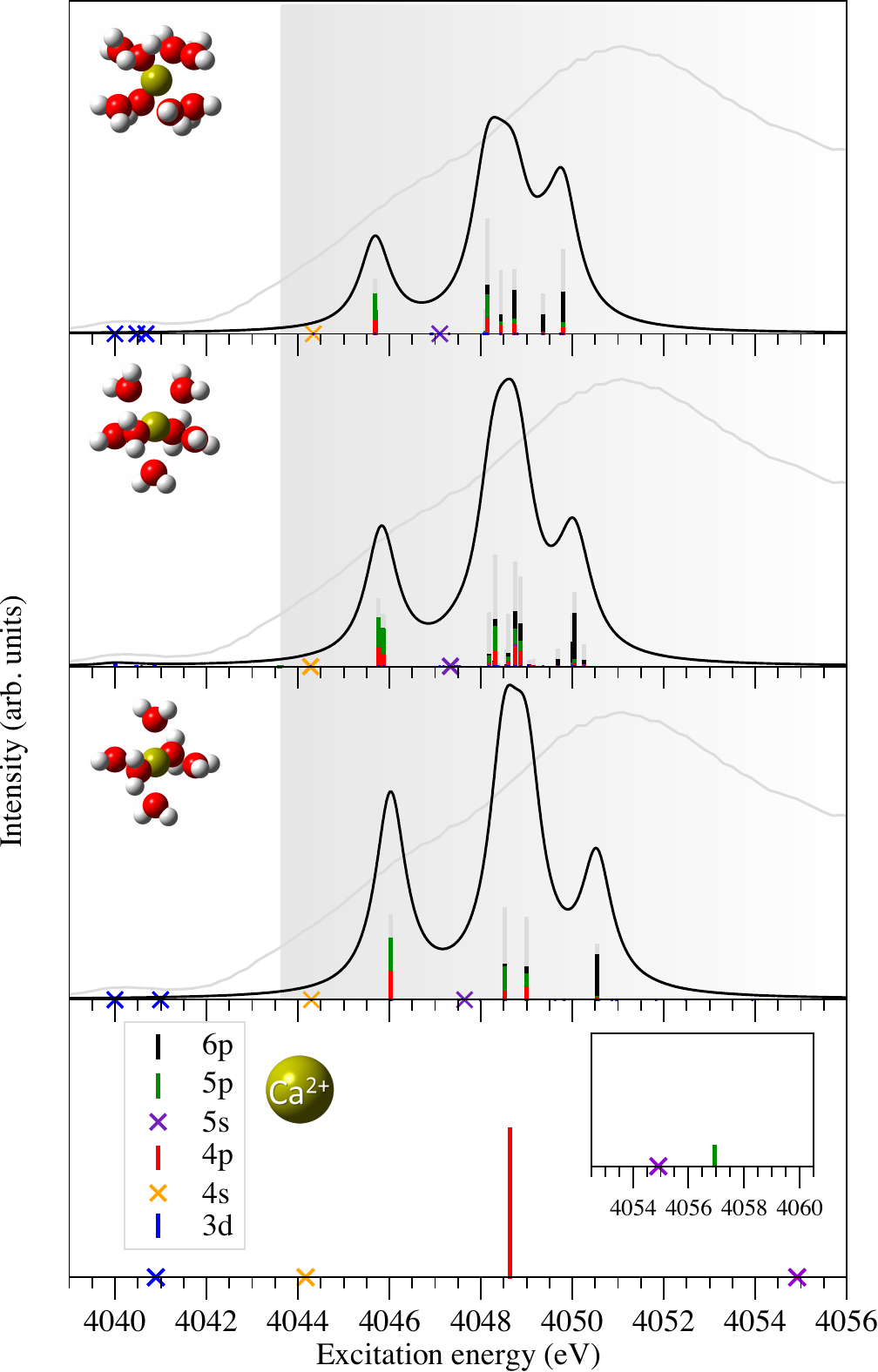}
    \caption{Theoretical K-shell core excited states computed with the CVS-ADC(2)-x method. Lowermost panel: lowest core excited states of the bare ion. Upper panels: lowest core excited states of [Ca(H$_2$O)$_6$]$^{2+}$, [Ca(H$_2$O)$_7$]$^{2+}$, [Ca(H$_2$O)$_8$]$^{2+}$. For more details, see text.}
    \label{fig:si1}
\end{figure}

The theoretical K-shell core excited states computed with the CVS-ADC(2)-x method, see main text, 
are shown in Fig.\ \ref{fig:si1}. The lowermost panel shows the lowest core excited states of the bare ion. The upper panels show from bottom to top the lowest core excited states of [Ca(H$_2$O)$_6$]$^{2+}$, [Ca(H$_2$O)$_7$]$^{2+}$, and [Ca(H$_2$O)$_8$]$^{2+}$. The black solid lines show Lorentzian convolutions of the stick spectra with FWHM of 0.81 eV\citep{krause79:329} representing the lifetime broadening. The colors in the stick spectrum 
represent the projections of the singly occupied natural orbitals (SONOs) of the core-excited 
6-, 7-, and 8-coordinated cluster on the basis of SONOs belonging to the 1s$^{-1}$3d/4s/4p/5s/5p/6p states 
of the isolated ion. Crosses indicate the position of the 3d and ns states whose intensity was calculated to be zero. The remaining contributions from higher lying atomic core excitations or from excitations 
to the solvent molecules are depicted as gray sticks. The theoretical spectra were shifted such 
that the Ca$^{2+}$(1s$^{-1}$3d) core excitation in the model clusters matches the experimentally
determined value. The experimental XAS spectrum is depicted as a grey line.

\begin{figure}[H]
\renewcommand{\thefigure}{SI\arabic{figure}}
    \centering
    \includegraphics[width=0.7\linewidth]{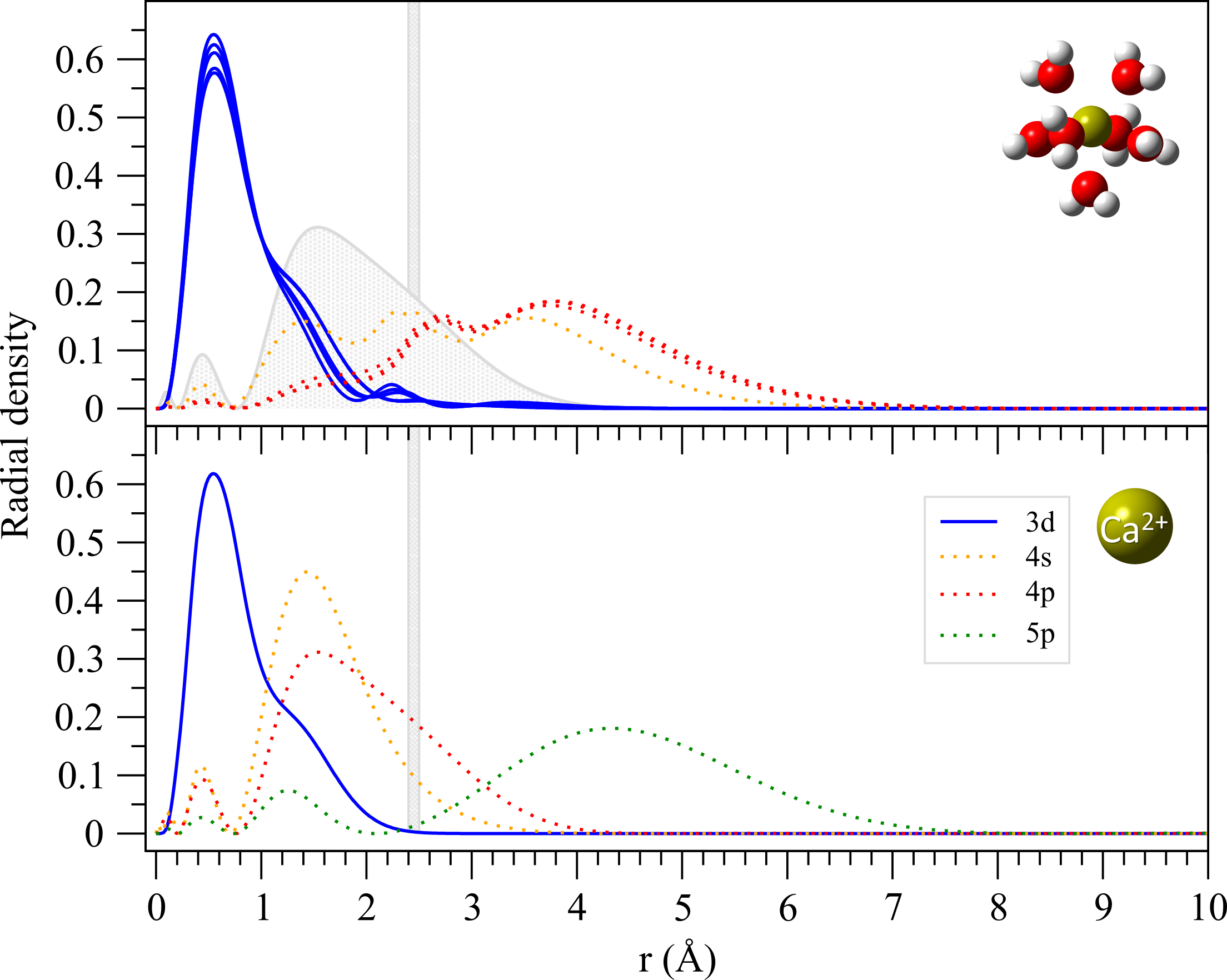}
    \caption{Radial density distributions of the SONOs occupied by the excited electron in the lowest 1s$^{-1}$3d/4s/4p/5p core excited states in the bare Ca$^{2+}$  ion (lower panel) and in the lowest core excited states of the 7-hydrated Ca$^{2+}$ ion correlating with the 1s$^{-1}$3d/4p bare-ion excitations. A gray vertical line designates the range of Ca$-$O distances in the first hydration shell obtained by geometry optimization of the [Ca(H$_2$O)$_{6-8}$]$^{2+}$ clusters. Upper panel: the gray area represents the radial density distribution of the SONO corresponding to the lowest dipole-allowed excitation (1s$^{-1}$4p) of the bare ion. }
    \label{fig:si2}
\end{figure}

\section{Post-collision interaction}
The PCI effect observed in the Auger spectra results from the Ca 1s photoelectron having a small kinetic
energy in the photon-energy region close to threshold. Thus, the photoelectron travels only a short distance before the Auger decay takes place (core-hole lifetime $\tau_{Ca 1s} \cong 0.81$ fs\citep{krause79:329}). For instance, without collision with the surrounding water molecules, an electron with 1 eV kinetic energy covers an average distance of about 5 \AA \ prior to the Auger decay. For comparison, the Ca$-$O bond distances of the first and second hydration shells are about 2.4 \AA \citep{jalilehvand2001:431,dangelo04:11857,tongraar10:10876} and about 4.55 
\AA  \citep{megyes04:7261}, respectively. On the contrary, the Auger electron has a high kinetic energy and quickly overtakes the photoelectron, which sees as a result a sudden change of the charge of the ionic core from Ca$^{3+}$ to Ca$^{4+}$. It is thus slowed down, whereas the Auger electron is accelerated leading to the positive shift seen in the left panel of Fig.\ \ref{fig:fg1} in the main text.

\section{Energy position of the Ca$_{aq}^{3+}$(1s$^{-1}$) $\rightarrow$ Ca$_{aq}^{4+}$(2p$^{-2}$($^1$S)) \linebreak Auger transition}

Since the energy position of the Ca$_{aq}^{3+}$(1s$^{-1}$) $\rightarrow$ Ca$_{aq}^{4+}$(2p$^{-2}$($^1$S)) Auger transition cannot be derived from the experiment, we estimated this quantity using the energy separation between the 2p$^{-2}$($^1$D) and 2p$^{-2}$($^1$S) states of different elements; this splitting increases with the atomic number of the atom or ion. An extrapolation of the experimental 2p$^{-2}$($^1$D) to 2p$^{-2}$($^1$S) splittings of Cl$_{aq}^-$ (9.0 eV)\citep{ceolin17:263003}, Ar (9.66 eV)\citep{puettner2020:052832}  and K$_{aq}^+
$ (10.4 eV)\citep{ceolin17:263003} results in a value of 11.1 eV for aqueous Ca$^{2+}$. This splitting is in line with the theoretical splitting of Ref.\citep{nishikida78:49}  estimated to 10.6 eV and 12.89 eV with two different approaches. In the present calculations we obtained, depending on the calculations, splittings between 13.0 and  13.5 eV, see below, which is also in agreement with the above mentioned values.

\section{Estimated energy position for a Cl to Ca charge-transfer state}

\begin{figure}
\renewcommand{\thefigure}{SI\arabic{figure}}
    \centering
    \includegraphics[width=0.6\linewidth]{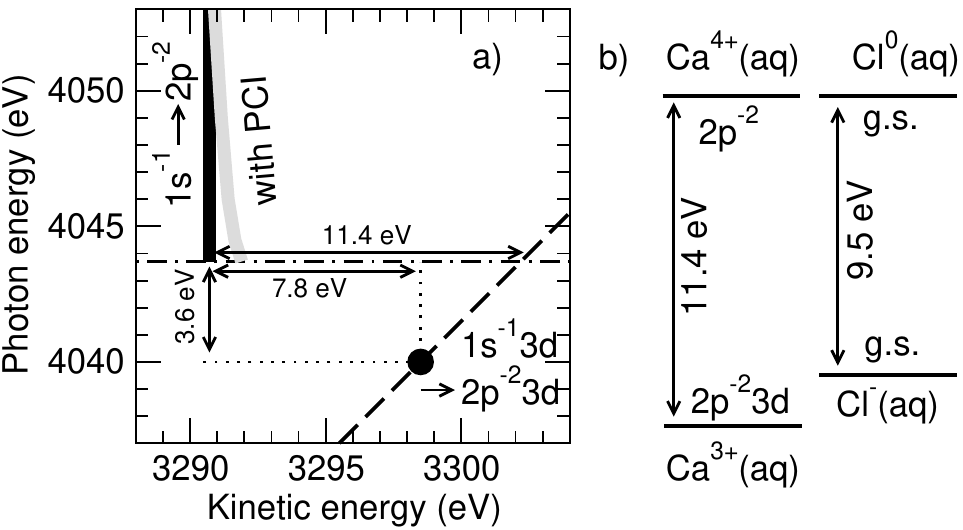}
    \caption{a) Schematic 2D-map formed by Auger spectra recorded at different photon energies with small step width. The x-axis represents the Auger kinetic energy and the y-axis the photon energies used for the excitation/ionization process. The dash-dotted line represents the ionization threshold, the black vertical line above the ionization threshold shows the Ca$_{aq}^{3+}$ (1s$^{-1}$) $\rightarrow$ Ca$_{aq}^{4+}$ (2p$^{-2}$) normal Auger lines without PCI effect and the filled circle the Ca$_{aq}^{2+}$ (1s$^{-1}$3d) $\rightarrow$ Ca$_{aq}^{3+}$ (2p$^{-2}$3d) resonant Auger transition. The diagonal dashed line indicates the linear dispersion for resonant Auger transitions to the final state Ca$_{aq}^{3+}$ (2p$^{-2}$3d) [peak F in the Auger spectra] and the lines with arrows at the ends indicate energy differences discussed in the text. In addition, the gray bend line in the upper part of the map shows the normal Auger contributions including the PCI effect. b) Energy scheme for the charge transfer from Cl$^{-}$ to the 3d orbital of the calcium ion with 2p$^{-2}$ core holes, which lead to an energy shift of $\Delta$E = $-$1.9 eV, see text.}
    \label{fig:si4}
\end{figure}

In the main text we suggest as a possible explanation for peak E a transfer of an electron from Cl$^-$   
in the second or third hydration shell to the 3d orbital of Ca$^{3+}$, i.e. Ca$_{aq}^{3+}$(1s$^{-1}$) + Cl$_{aq}^-$   $\rightarrow$ Ca$_{aq}^{3+}$(2p$^{-2}$3d) + Cl$_{aq}^0$ . This process can be suppressed by the presence of a slow photoelectron, see main text. Here we present a simple model which describes reasonably the energy position of the suggested Auger transfer. In detail, the process can be understood by an ionization of Cl$^-$ from the outermost valence shell and a subsequent attachment of the electron in the 3d orbital of Ca$^{3+}$, which accompanies the 1s$^{-1}$  $\rightarrow$ 2p$^{-2}$  Auger decay. In a simple model shown in Fig.\ \ref{fig:si4}b the Auger energy is shifted by the difference of the ionization energy of Cl$_{aq}^-$  and the term value of the 3d electron in Ca$_{aq}^{3+}$. According to the literature, the value of the ionization energy of Cl$_{aq}^-$ is between 8.7 and 9.6 eV \citep{winter2005:7203,kurahashi2014:174506}; in Figure \ref{fig:si4}b we use a value of 9.5 eV. The term value of the 3d electron, i.e.\ the energy  positions of the states Ca$_{aq}^{2+}$(1s$^{-1}$3d) and Ca$_{aq}^{3+}$(2p$^{-2}$3d) relative to the corresponding ionization threshols can be obtained from the present X-ray absorption, X-ray photoelectron and Auger spectra. Note that these term values depend on the surrounding water, i.e., the number of water molecules in the first solvation shell as well as their special arrangement.  

To derive the term value for the 3d electron, we focus on the Ca$_{aq}^{2+}$ (1s$^{-1}$3d) $\rightarrow$
Ca$_{aq}^{3+}$ (2p$^{-2}$3d) resonant Auger decay, which causes peak F in the Auger spectra. The term value
of the initial state amounts to $\cong$3.6 eV, see the discussion of the X-ray absorption spectrum in the main
text. From the kinetic energy of the resonant Auger decay, peak F, which is $\cong$7.8 eV larger than that of the
corresponding main line (peak A) without PCI shift, see Fig.\ \ref{fig:fg1} of the main text, we obtain a term value of
$\cong$11.4 eV for Ca$_{aq}^{3+}$ (2p$^{-2}$3d) relative to the Ca$_{aq}^{4+}$ (2p$^{-2}$) double ionization threshold, see Fig.\ \ref{fig:si4}a. From the ionization energy of  Cl$_{aq}^-$ and the term value of the 3d orbital we obtain in our simple model an energy release of about 2 eV, i.e.\ a shift of the corresponding CT Auger line by this value to higher kinetic energies as compared to the main line (structure A). This value agrees quite well with the experimental splitting between structure A and E of about 4 eV. 

\section{Theoretical KLL Auger spectra}

Simulations of the KLL Auger spectrum of solvated Ca$^{2+}$ were performed on a cluster reduced to a Ca$^{2+}$ ion surrounded by two neutral water molecules forming a linear conformation within the D$_{2h}$ symmetry. For calcium  the Pople Valence Triple zeta basis set (6-311G*) augmented with five d functions of exponents 40/20/0.26/0.08/0.026 is used. The chosen basis set takes into account the diffuse character of the electronic transitions of interest (H$_2$O $\rightarrow$ 4s, 4p, 5s, etc) and the fast Auger electron ejected in the continuum with a kinetic energy around 3300 eV as recorded in the experiment. 
In order to simulate the KLL Auger spectrum, a two-step configuration interaction (CI) calculation was performed. The first step used a CI space reduced to spin-singlet 2p$^{-2}$ configurations in order to estimate the 2p$^{-2}$($^1$D)$-$2p$^{-2}$($^1$S) energy gap. With this space, the numerical gap is theoretically estimated to be about 13.5 eV. A larger CI (SDT) calculation reduced the gap to 13.0 eV, which, as discussed above, is in good comparison with the literature. In order to account for the experimental 2p$^{-2}$($^1$D)$-$2p$^{-2}$($^1$S) splitting, a constant offset was applied to the 2p$^{-2}$($^1$S) calculated Auger peaks.

\begin{figure}
\renewcommand{\thefigure}{SI\arabic{figure}}
    \centering
    \includegraphics[width=0.7\linewidth]{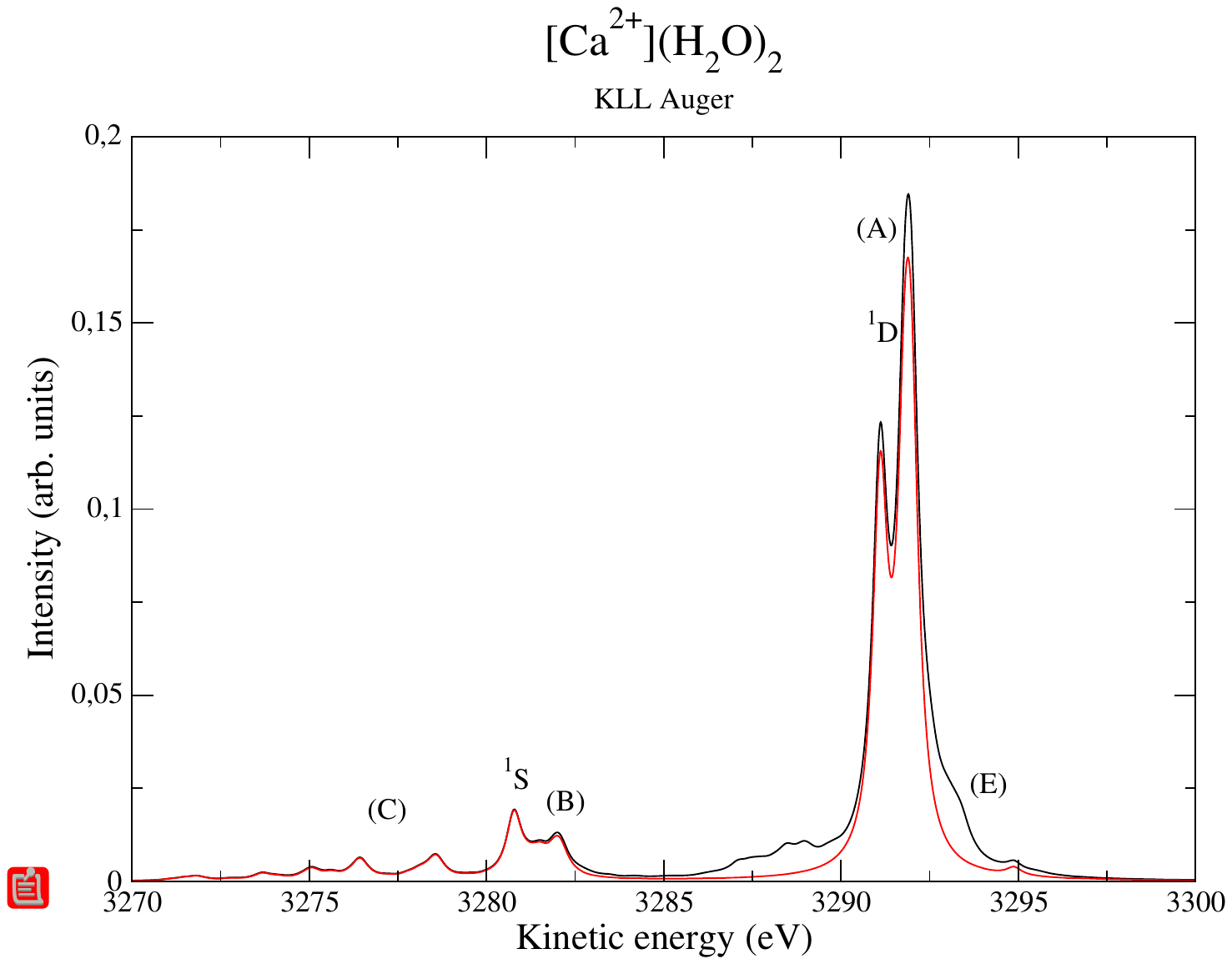}
    \caption{Theoretical KLL Auger spectra of the [Ca(H$_2$O)$_2$]$^{2+}$ microsolvated cluster. Red line: KLL Auger spectrum of the Ca 1s$^{-1}$ core-ionized state, black line: KLL Auger spectrum of the Ca 1s$^{-1}$ state and the photoelectron satellites up to 15 eV above the main line.}
    \label{fig:si_auger_theory}
\end{figure}

The second step considers separately spin doublet 2p$^{-2}\varepsilon$d and 2p$^{-2}\varepsilon$s configurations. These configurations include the states 2p$^{-2}$($^1$D)$\varepsilon$d($^2$S) and 2p$^{-2}$($^1$S)$\varepsilon$s($^2$S), respectively, which are the only ones leading to  non-vanishing Auger intensities. In the CI active space we have considered between 3 and 5 unpaired electrons.  

The Auger intensities were computed evaluating the squares of Coulomb-exchange integrals considering spin conditions between the doublet core-ionized Ca$^{3+}$(1s$^{-1}$) intermediate state and the Ca$^{4+}$(2p$^{-2}$) final states. We considered here an Auger process consisting of core ionization (1s $\rightarrow \psi_{\varepsilon p}$) 
and decay ($\phi_{2p} \rightarrow 1s$), ($\phi_{2p'} \rightarrow \psi_{\varepsilon s,d}$)
steps. The Coulomb matrix element for the Auger decay is given by the following expression
\begin{equation*}
    \langle \psi_{\varepsilon} \phi_{1s}|\phi_{2p} \phi_{2p'} \rangle = \int \psi_{\varepsilon} (\vec{r_1}) \phi_{2p'}
    (\vec{r_1}) \frac{1}{r_{12}} \phi_{1s} (\vec{r_2})  \phi_{2p} (\vec{r_2}) d\vec{r_1} d\vec{r_2}.
\end{equation*}
The simulated Auger spectrum is shown in Fig.\ \ref{fig:si_auger_theory}. The red line shows the theoretical Auger spectrum originating from the decay of the 1s$^{-1}$ core-ionized state, whereas the black line corresponds to Auger decays of the 1s$^{-1}$ state and of the photoionization satellite states found
in the theoretical XPS spectrum (see Fig.\ 1 of the main text) up to 15 eV above the main line. 

Our theoretical calculations show that region A consists of two main lines, both corresponding to final configurations associated with the 2p$^{-2}$($^1$D) atomic state. Analysis of the CI states indicates that these lines represent a mixture of the $^1$D atomic configuration with exications from water to the 3d, 4s, and 4p orbitals of the metal ion. The splitting of these lines can be attributed to a molecular field effect, related to the different localization of the excited electron in the Ca 4p orbital (4p$_{x}$ or 4p$_z$), which is oriented either parallel or perpendicular to the Ca$-$H$_2$O chemical bond. Additionally, the splitting arises from the various spin functions contributing to 
the CI states. Note that the ligand-field splitting obtained for  the geometry of [Ca(H$_2$O)$_2$]$^{2+}$ used in the calculations is expected to be much larger than that of the experiment. This is due to the fact that in the experiment Ca$^{2+}$ is surrounded by many water molecules which form an almost isotropic environment, see Fig.\ \ref{fig:si1}.

At kinetic energies lower by 10 to 15 eV than the main peak A, the experimental Auger spectrum reveals the presence of two broad structures, labeled B and C. The theoretical KLL Auger spectra confirm that these peaks correspond to a superposition of Auger transitions to the atomic 2p$^{-2}$($^1$S) configuration, along with transitions to other 2p$^{-2}$ configurations involving charge transfer from water to the nd orbitals of the metal ion. This is consistent with the calculated Auger final states (Fig.\ 4 in the main text). Additionally, the calculations suggest that region C primarily represents single or multiple excited configurations involving interactions between water (w,w') and calcium energy levels, such as ww’ $\rightarrow$ nd/4s/4p.


Despite the limited representation of the Ca$^{3+}$  1s$^{-1}$  satellite energy region in our Ca(H$_2$O)$_2$ cluster model and the 
mismatch between the theoretical and experimental energy positions of peak E, including satellite states in the 
Auger simulation significantly increases the signal intensity in regions E and A.
These bands result from the Auger decays of CT satellite states accompanying the Ca 1s ionization and including excitations from water to Ca 3d orbitals, located at binding energies 10$-$15 eV above the main peak (see Fig.\ \ref{fig:fg1} in the main text). They contribute to the Auger decay in region E through a 
process, leading to the population of final states similar to those populated during the charge-transfer process. 
Auger decay of the Ca$^{3+}$  1s$^{-1}$ photoelectron satellite states is therefore one of the contributions to feature E, however, only at photon energies above 4054 eV.

\section{On photoelectron satellites }

Shake-up satellites is a common term used to refer to photoelectron satellites. However, 
photoelectron satellites in an isolated system are generally due to electron correlation 
\citep{yarzhemsky2016:063406} and there are different
mechanisms to populate such satellite states. Since the shake process is only one of them 
we avoid in this publication the term shake-up satellites. As discussed in the main text, in a surrounding like water inelastic photoelectron scattering (IPES) can lead to satellites. For isolated systems
there are different intrinsic processes which lead to the formation of satellite structures, 
namely shake, knock, and charge transfer which shall be discusses shortly.

The shake process is based on the sudden approximation, i.e.\ the outgoing photoelectron possesses a 
high kinetic energy so that it leaves the system quickly without direct interaction with the other electrons
\citep{dyall1983:3137}.
In the following we discuss the shake process using an atom as an example, although the arguments are also valid
for molecules or condensed matter. In the atom, the photoionization leads to the fact that the remaining electrons see a different charge before and after ionization. As a result, the radial wavefunctions of the orbitals before and after ionization are also different.  Because of this, the orbitals prior to ionization, nl, are not orthogonal to 
the orbitals subsequent to ionization, n$^{\prime}$l$^{\prime}$, i.e.\ $\langle$nl$|$n$^{\prime}$l$^{\prime}\rangle \ne 
\delta_{nn^{\prime}} \delta_{ll^{\prime}}$.
The shake process requires the same symmetry of the two orbitals involved, which in case of an atom is given by the angular 
momentum l, i.e.\ l = l$^{\prime}$ is required; such processes are also called monopole transitions. The intensity of the shake satellites is described by $|\langle$nl$|$n$^{\prime}$l$\rangle|^2$.

In contrast to the shake processes it is also possible that the outgoing electron interacts
directly with another electron in the system\citep{yarzhemsky2016:063406,defanis2004:265}. Such a process occurs in particular if the outgoing electron is
slow since this leads to a long Coulomb interaction with the other electrons and therefore to a higher probability
for a transfer of energy and angular momentum. Such a direct interaction between the outgoing electron and the
electron promoted from one orbital to the other is called a knock process. Moreover, the interaction can also
lead to an exchange of angular momentum and a change of the orbital symmetry (e.g.\ in an atom from a np orbital to a md orbital); such processes are labeled non-monopole transitions. 

Charge transfer satellites are a complex matter so that we present here just some 
principle ideas. A more comprehensive discussion of theses satellites is presented in\citep{huefner2003}.
Generally, charge transfer states require two systems (here Ca$^{2+}$ and H$_2$O), which interact
only weakly, i.e.\ show only a weak hybridization. The ionization process (here of the calcium 
1s level) can lead to a total or partial transfer of an electron from the non-ionized system (here water) to the ionized system. This behavior can be understood based on a molecular-orbital model, which we will explain based on the present system consisting of calcium and water. Without interaction between
calcium and water the electron has to be located on one system, namely on calcium or on water. This gives us before ionization the two states $|3p^6 W \rangle$, i.e.\ Ca$^{2+}$ + H$_2$O, and $|3p^6 3d^1 W^- \rangle$, i.e.\ Ca$^{+}$ + H$_2$O$^{+}$, where $W$
stays for the water molecule and $3p^6 3d^n$ for the occupation of the calcium valence orbitals. 
And after ionization we obtain the states $|1s^{-1} 3p^6 W \rangle$, i.e.\ Ca$^{3+}$ + H$_2$O, and $|1s^{-1} 3p^6 3d^1 W^- \rangle$, i.e.\ Ca$^{2+}$ + H$_2$O$^{+}$. If we now allow interaction 
(hybridization) between water and calcium, we obtain for the ground state $\alpha | 3p^6 W \rangle + \beta|3p^6 3d^1 W^- \rangle$ with $\alpha \gg \beta$ since the charge-transfer state
Ca$^{+}$ + H$_2$O$^{+}$ is significantly higher in energy. In addition, we obtain two core-ionized
states, namely $|1\rangle = \alpha^{\prime}|1s^{-1} 3p^6 W \rangle + \beta^{\prime}|1s^{-1}3p^6 3d^1 W^- \rangle$ and $|2\rangle = \beta^{\prime}|1s^{-1} 3p^6 W \rangle - \alpha^{\prime}|1s^{-1}3p^6 3d^1 W^- \rangle$, which both can be populated via 1s ionization of the ground state. Depending
on the values for $\alpha^{\prime}$ and $\beta^{\prime}$ in one of the states the electron is 
predominantly on the water site (normal state) or in the Ca 3d orbital (charge transfer state).
Note that in general $\alpha \ne \alpha^{\prime}$ and $\beta \ne \beta^{\prime}$ since the energy
difference between the states $| 3p^6 W \rangle$ and $|3p^6 3d^1 W^- \rangle$ is different
from that of $| 1s^{-1}3p^6 W \rangle$ and $|1s^{-1}3p^6 3d^1 W^- \rangle$.

The lower the symmetry of the system, the more difficult it becomes to unambiguously characterize
the process which is responsible for the formation of photoelectron satellites, i.e.\ different processes can 
lead to the same satellite state visible in the photoelectron spectrum. Nevertheless, a general characterization of
all photoelectron satellites as shake structures oversimplifies their wealth of formation processes.
Finally we want to point out that all described processes can also lead to satellite structures in the Auger spectra.


\bibliography{cacl2_bib}